\documentclass[aps,pre,showpacs,floats,twocolumn,superscriptaddress,floatfix,english]{revtex4-1}
\usepackage[T1]{fontenc}
\usepackage[latin9]{inputenc}
\usepackage{color}
\usepackage{amsmath}
\usepackage{graphicx}

\makeatletter

\usepackage{array}
\usepackage{multirow}
\usepackage{setspace}

\providecommand{\tabularnewline}{\\}
\usepackage{multirow}
\usepackage{caption}
\usepackage{color}
\usepackage{epstopdf}
\usepackage{epsfig}
\usepackage[normalem]{ulem}
\usepackage{babel}
\usepackage{float}
\newcommand{\nordita}{Nordita, KTH Royal Institute of Technology and Stockholm University, Stockholm 10691, Sweden}

\makeatother

\usepackage{babel}
\begin{document}
\title{Melting driven by rotating Rayleigh-B\'enard convection}
\author{S. Ravichandran}
\affiliation{\nordita}
\author{J. S. Wettlaufer}
\affiliation{\nordita}
\affiliation{Yale University, New Haven, Connecticut 06520-8109, USA}
\makeatother

\begin{abstract}
We study numerically the melting of a horizontal layer of a pure solid above a
convecting layer of its fluid rotating about the vertical axis.  In the rotating regime studied here, with Rayleigh numbers of order $10^{7}$, 
convection takes the form of columnar vortices, the number and size of which depend upon the Ekman and Prandtl numbers, as well
as the geometry--periodic or confined.  As the Ekman and Rayleigh numbers vary, the number and average area of vortices vary in inverse
proportion, becoming thinner and more numerous with decreasing Ekman number.  The vortices transport heat to the phase boundary thereby
controlling its morphology, characterized by the number and size of the voids formed in the solid, and the overall melt rate, which increases 
when the lower boundary is governed by a no-slip rather than a stress-free velocity boundary condition.  Moreover, the
number and size of voids formed are relatively insensitive to the Stefan number, here inversely proportional to the latent heat of fusion. For small values of the Stefan number, the convection in the fluid reaches a slowly evolving geostrophic state wherein columnar vortices transport nearly all the heat from the lower boundary to melt the solid at an approximately constant rate.
In this quasi-steady state, we find that the Nusselt number, characterizing the heat flux, {co-varies with} the interfacial roughness, {for all} the flow parameters and Stefan numbers {considered here}. This confluence of processes should influence the treatment
of moving boundary problems, particularly those in astrophysical and geophysical problems where rotational effects
are important.
\end{abstract}
\maketitle

\section{Introduction \label{sec:Introduction}}

The coupling between a solid and the liquid from which it forms controls
the long term fate of both phases. Through deliberate manipulation
of the flow of the nutrient phase, engineers aim to control the character
of a solidified material \citep{DavisBook}. When the heat transport
required for solidification occurs through diffusion, initially planar
phase-boundaries remain planar. But the presence of convection invariably leads to
non-planar interfaces. The uncontrolled interplay of convection, rotation,
and phase change determines the dynamics of many geophysical and astrophysical
systems. Indeed, such processes operate from Earth's core to the principal
components of the cryosphere \citep[e.g.,][]{HEH:1990,worster2000solidification}.
In astrophysics, they underlie planet formation \citep[e.g.,][]{armitage2020astrophysics},
wherein for example the proto-Earth was believed to rotate about ten
times faster than today \citep[e.g.,][]{Cuk:2012}, and the growth
of neutron star crusts \citep[e.g.][]{baym2018hadrons}, amongst many
other phenomena. The confluence of dynamic and thermodynamic processes
in such systems is highly complex and involves multiple timescales,
components and phases.

Here, we study a simplified system of a single-component rotating
phase boundary heated from below. The associated rotation-influenced
convection brings heat to the solid upper boundary, controlling the morphology of the melting solid.

A non-rotating layer of fluid heated from below begins convecting when the thermal buoyancy overcomes the viscous and thermal dissipation effects
that suppress vertical motions. This balance is characterized by the dimensionless Rayleigh number
\begin{equation}
Ra=\frac{g\alpha\Delta Th^{3}}{\nu\kappa},\label{eq:Rayleigh_basic}
\end{equation}
where $g$ is the acceleration due to gravity; $\alpha$, $\nu$ and
$\kappa$ are the coefficient of thermal expansion, the viscosity
and the thermal diffusivity of the fluid; and $h$ is the depth of
the fluid layer across which a temperature difference $\Delta T$
is imposed.  Convective motions begin when $Ra$ exceeds a critical value $Ra_{c} = \mathcal{O}\left(10^{3}\right)$, the 
prefactor depending on the boundary conditions \citep[e.g.,][]{chandrasekhar1961}.

In direct analogy with stratification in non-rotating systems, rotation suppresses vertical motions due to buoyancy \citep{GV:ARFM}.
Therefore, the critical Rayleigh number above which convection occurs is a function of the rotation rate of the system \citep{Chandrasekhar1953}.
The Ekman number is the relevant nondimensional rotation rate and is
\begin{equation}
E=\frac{\nu}{2\Omega h^{2}},\label{eq:Ekman_basic}
\end{equation}
where $\Omega$ is the angular velocity of the system.  Thus, rapidly rotating systems are characterized by small
$E$.  Whereas in non-rotating convection a given set of boundary conditions determines the single value of $Ra_{c}$, 
in rotating convection $Ra$ is an increasing function of $E^{-1}$, where both the functional form and numerical factors depend on the
boundary conditions of the problem.

If the horizontal directions are assumed to be periodic, the onset
of convection occurs above $Ra_{c}\sim E^{-4/3}$. For one free-slip
one no-slip boundary each (and periodic boundary conditions in the
horizontal), in the limit of large $E^{-1}$ \citep{Chandrasekhar1953},
$Ra_{c}$ is 
\begin{equation}
Ra_{c}^{\text{bulk}}=2.39E^{-4/3}.\label{eq:Ra_cr_bulk}
\end{equation}
If the horizontal directions are bounded by walls, the critical Rayleigh
number for the so-called `wall-mode' \citep{Zhong:1991, Ecke:1992}
is, in the limit of large $E^{-1}$, given by \citep{Herrmann1993}
\begin{equation}
Ra_{c}^{\text{wall}}|_{E^{-1}\rightarrow\infty}=\pi^{2}(6\sqrt{3})^{1/2}E^{-1}<Ra_{c}^{\text{bulk}}.\label{eq:Ra_cr_wall}
\end{equation}
In a rotating system bounded laterally by walls, flow is absent for
$Ra<Ra_{c}^{\text{wall}}$. The flow structures that appear for $Ra>Ra_{c}^{\text{wall}}$
take the form of a peripheral streaming current adjacent to the walls,
with alternating bands of up- and down-welling flow. While the flow
in them is still cyclonic, these patterns precess about the axis of
rotation in a retrograde direction \citep{HornSchmid2017,Favier2020wallmodes,DeWit2020,Zhang2020},
even when there are severe obstacles in the way \citep{Favier2020wallmodes}.
These wall-modes persist even when the bulk of the flow begins to
convect, and they underlie an observed mismatch between theoretical and numerical predictions
of heat transport and laboratory observations at large $Ra$ \citep{DeWit2020}.

When $Ra>Ra_c^{\text{bulk}}$, convection begins throughout the fluid. For $Ra\lesssim 10 Ra_{c}^{\text{bulk}}$,
flow occurs along columnar (Taylor) vortices that span the depth of
the fluid \citep{boubnov1986,boubnov1990,Zhong:1991,king2009boundary,Aurnou2015}.
These vortices are predominantly cyclonic near the upper and lower
boundaries, with equal numbers of cyclonic and anticyclonic vortices
in the interior \citep{VorobieffEcke1998,zhong1991asymmetric,Kunnen2010,boubnov1986},
thereby transporting heat from the boundaries \citep{Sakai1997}. For $Ra>10 Ra_{c}^{\text{bulk}},$
the columnar vortices become plume-like and lose their vertical
alignment with the axis of rotation. The highest Rayleigh numbers
achieved in our simulations are in this regime. For sufficiently large
$Ra$ (and sufficiently large $E^{-1}$), a state of `geostrophic turbulence'
sets in \citep{boubnov1990,King2012,Shi2020}, a computationally challenging
regime to study.

The nature of rotating convection and the rate of heat transport are controlled by the combination of $E$, $Ra$ and $Pr$, 
and thus so too will be the {melt rate and} patterns of an adjacent phase boundary, such as we study here.
{While varying the dimensionless latent heat, or Stefan number, is expected to affect the overall rate of phase change, the effects
on the interfacial patterns that form are more subtle, which largely reflect the nature of the transport properties of the bulk flow.} 
This confluence of effects form the core of our study.

The rest of the paper is organized as follows. We describe the structure
of the problem in \S \ref{sec:Setup}, providing details of the phase
change treatment used; the approximations made; the relevant physical
scales and the nondimensionalization; the boundary and initial conditions;
and the numerical algorithm used to solve the governing equations.
In \S \ref{sec:Results}, we discuss the effects of the control parameters on the phase boundary morphology, which is 
dominated by rotation.  We obtain the melt rates and their associated Nusselt number dependencies.  
Additionally, we discuss how the dynamics change if the system is periodic in the horizontal,
if the lower boundary is one of no-slip, and when the solid
has a thermal diffusivity different from the liquid. We conclude
with some ideas for future work.

\section{Structure of the Problem \label{sec:Setup}}

\begin{figure}
\includegraphics[width=0.48\columnwidth]{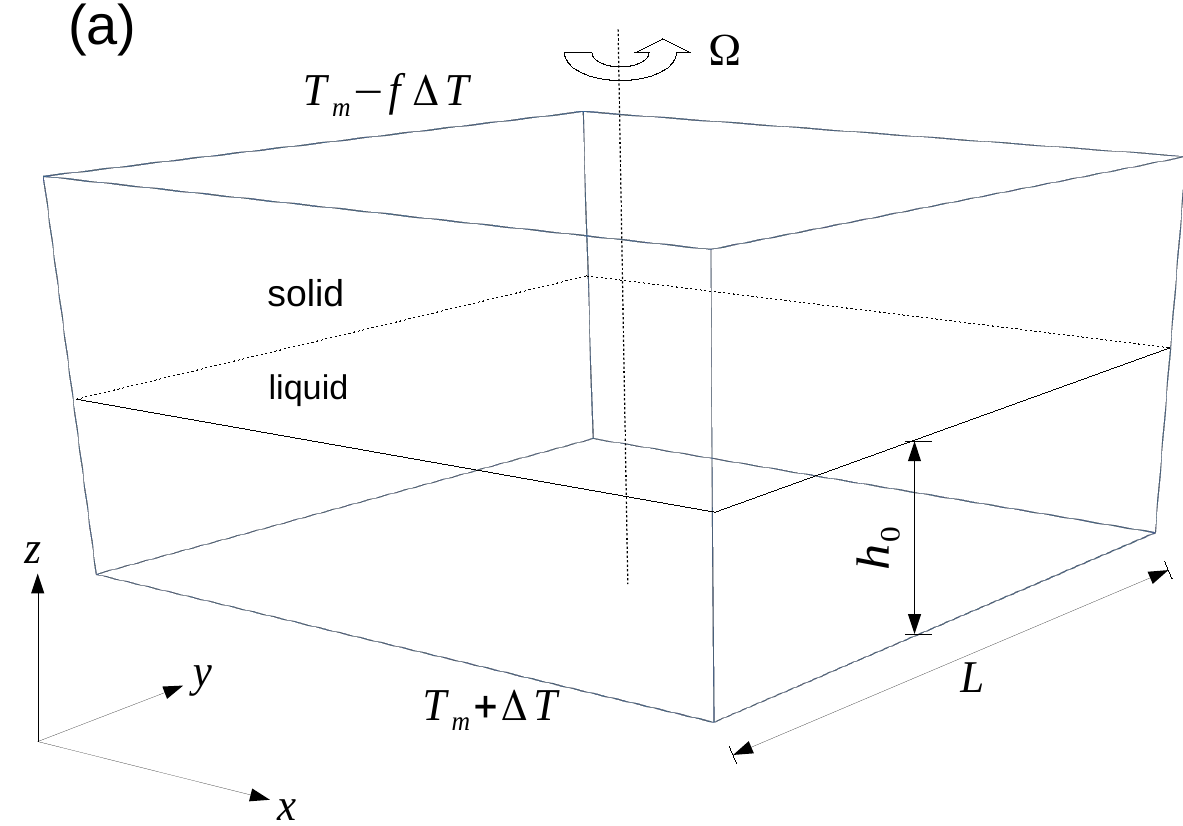}
\includegraphics[width=0.48\columnwidth]{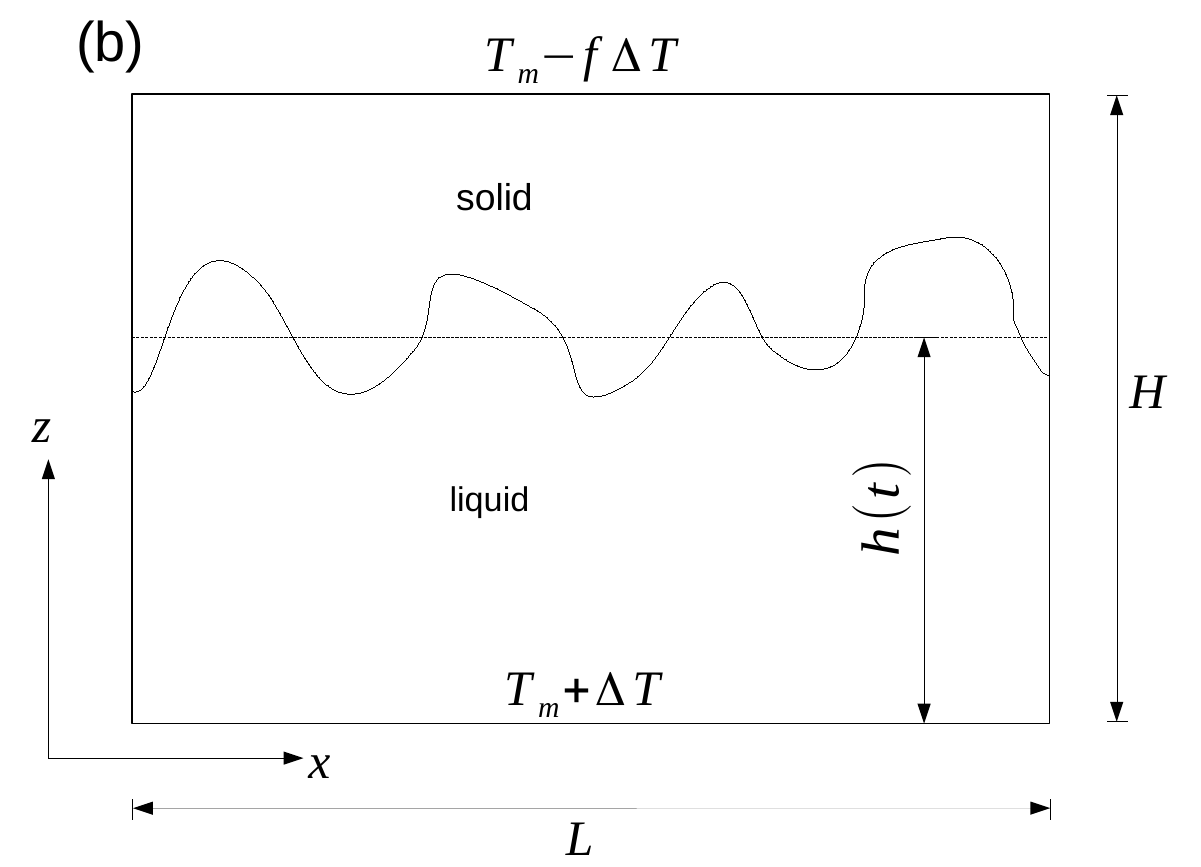}
\caption{\label{fig:schematic} {(a) A schematic of the geometry used, with the coordinate directions and dimensions marked. The initial liquid height is $h(t=0) = h_0$. (b)} Vertical cross-section of the geometry considered { at $t>0$}. The system rotates about the $z$ axis, and gravity is in the $-z$ direction. $T_m$ is the melting temperature of the pure substance, and the lower boundary is at temperature $T_m + \Delta T$.  The effective Rayleigh and Ekman numbers are defined based on the {horizontally averaged} fluid height $h(t)$, while the reference values are defined based on $H/2$ where $H$ is the height of the solid+liquid system.}
\end{figure}

Our study geometry is a box of dimensions $L\times L\times H$,
with gravity $g$ in the $-z$ direction, and rotating about the $+z$
axis with an angular velocity $\Omega$, shown schematically in Fig.
\ref{fig:schematic}. The aspect ratio of the simulation domain is
$L/H=2$. The mean height of the liquid layer at time
$t$ is $h\left(t\right)$, with $h\left(t=0\right)=h_{0}$. We {use
the domain half-height as our length scale (see \S \ref{subsec:Governing-equations} below),
and define the aspect ratio as $A=2L/H$}. The system is heated from below by imposing 
a constant temperature difference between the lower and upper boundaries, thereby thereby melting the solid. 
{As described in \S \ref{subsec:ICs_BCs}, the majority of our results are obtained with the entire solid at the melting temperature, so that there is no heat conduction through the solid.}

\subsection{Enthalpy Method \label{subsec:Enthalpy-method}}

We employ a mixture theory approach to tracking the solid region,
such that a solid fraction variable $\chi$ varies from $0$ in the
liquid state to $1$ in the solid state. 
The densities of the solid and liquid phases
are $\rho_{s}$ and $\rho_{l}$ respectively; their heat capacities
are $C_{s}$ and $C_{l}$ respectively; and the latent heat of fusion
is $\lambda$. Here, for simplicity, we only consider
the case where the solid and liquid have the same densities and 
\begin{align}
\rho_{s} & =\rho_{l}\left(=\rho\right)\label{eq:approx_const_density}\\
C_{s} & =C_{l}\left(=C_{p}\right)\label{eq:approx_const_Cp}
\end{align}
with $\rho$ and $C_{p}$ being constants. The solid
and liquid enthalpies are 
\begin{align}
{\cal H}_{s} & =\rho C_{p}T,\text{ and}\label{eq:dim_solid_enthalpy}\\
{\cal H}_{l} & =\rho C_{p}T+\rho\lambda,\label{eq:dim_liq_enthalpy}
\end{align}
respectively. The enthalpy of the solid phase at the melting temperature
$T_{m}$ is ${\cal H}_{0}=\rho C_{p}T_{m}$, and that of a mixture
of solid and liquid phases with solid volume fraction $\chi$ is given
by 
\begin{align}
{\cal H} & =\chi\rho C_{p}T+\left(1-\chi\right)\rho\left[C_{p}T+\lambda\right].\label{eq:dim_net_enthalpy}\\
 & =\rho C_{p}T+\left(1-\chi\right)\rho\lambda\nonumber
\end{align}
We nondimensionalize the enthalpy as 
\begin{equation}
\phi=\frac{{\cal H}-{\cal H}_{0}}{\rho C_{p}\Delta T}=\frac{T-T_{m}}{\Delta T}+\frac{\lambda}{C_{p}\Delta T}\left(1-\chi\right),\label{eq:defn_phi}
\end{equation}
where $\Delta T$ is the difference between the temperature of the
lower boundary and the melting temperature. Thus, if 
\begin{eqnarray}
\theta & = & \frac{T-T_{m}}{\Delta T}\label{eq:theta}
\end{eqnarray}
is defined to be the nondimensional temperature, and 
\begin{equation}
St=C_{p}\Delta T/\lambda\label{eq:Stefan}
\end{equation}
is the Stefan number (often also defined as the inverse of this) then we have 
\begin{equation}
\phi=\theta+St^{-1}\left(1-\chi\right). 
\label{eq:phi}
\end{equation}
We note that in the purely solid state
$\chi=1$ and ${\theta\leq0},$ so that $\phi\leq0$. The equation of
state \eqref{eq:phi} can be inverted to give the solid fraction in
terms of the enthalpy as 
\begin{equation}
\chi=1-\textrm{max}\left[0,\textrm{min}\left(1,St\ \phi\right)\right],\label{eq:chi}
\end{equation}
and hence the temperature follows as 
\begin{equation}
\theta=\phi-St^{-1}\left(1-\chi\right).\label{eq:EOS_rev}
\end{equation}
Thus, in a pure solid, $\chi=1$, $\theta=\phi$; in a pure liquid,
$\chi=0$, $\theta=\phi-St^{-1}$; in the mixed phase, $0<\chi<1$
and $\theta = 0$, by definition.
In the vicinity of the phase boundary $\chi$ must change from 0 to 1 over a very thin region
\citep[see e.g.,][]{RabbanipourEsfahani2018,Favier2019}, 
which is a requirement that our simulations obey. 
The normal motion of the
phase boundary, $u_{m}$, is determined by the interphase difference
between heat fluxes, and the Stefan condition in dimensional variables
is 
\begin{equation}
\rho\lambda u_{m}=k_{s}\left(\nabla T\right)_{s}-k_{l}\left(\nabla T\right)_{l},\label{eq:meltvel}
\end{equation}
where $\left(\nabla T\right)_{s}$ and $\left(\nabla T\right)_{l}$
are the temperature gradients in the solid and the liquid on either
side of the phase boundary; and $k_{s}$ and $k_{l}$ are the thermal
conductivities in the solid and liquid respectively.

\subsection{Governing Equations \label{subsec:Governing-equations}}

The equations of motion that govern the evolution of the velocity
$\boldsymbol{u}$, and the enthalpy $\phi$, defined in Eq. \eqref{eq:phi},
are as follows. We study the rotating Oberbeck-Boussinesq equations
with the assumptions in Eqs. \eqref{eq:approx_const_density} and
\eqref{eq:approx_const_Cp}, which are
\begin{widetext}
\begin{eqnarray}
\frac{D\boldsymbol{u}}{Dt} & = & -\frac{\nabla p}{\rho}+\nu\nabla^{2}\boldsymbol{u}+g\alpha\mathbf{e}_{z}\left(T-T_{m}\right)-2\Omega\mathbf{e}_{z}\times\boldsymbol{u},\label{eq:dim_momentum}\\
\frac{D\theta}{Dt} & = & \nabla\cdot\left(\kappa\nabla\theta\right),\text{ and}\label{eq:dim_enthalpy}\\
\nabla\cdot\boldsymbol{u} & = & 0\label{eq:dim_continuity}
\end{eqnarray}
\end{widetext}
where $\alpha$ is the coefficient of thermal expansion, $\nu$ is
the kinematic viscosity of the fluid, and $\kappa=\chi\kappa_{s}+\left(1-\chi\right)\kappa_{l}$
is the local thermal diffusivity. These equations are nondimensionalised
using the temperature scale $\Delta T$ from Eq. \eqref{eq:phi}, and
{the length scale $H/2$, where $H$ is the height of the domain}, giving
a buoyancy velocity {$U_{b}=\left(g\alpha\Delta T H/2\right)^{1/2}$}.
Using these scales, the dimensionless equations of motion become 
\begin{widetext}
\begin{eqnarray}
\frac{D\boldsymbol{u}}{Dt} & = & -\nabla p+\left(\frac{Pr}{Ra}\right)^{1/2}\nabla^{2}\boldsymbol{u}+\mathbf{e}_{z}\theta-{Ro_c^{-1}}\mathbf{e}_{z}\times\boldsymbol{u},\label{eq:momentum}\\
\frac{D\theta}{Dt} & = & \left(\frac{1}{Ra Pr}\right)^{1/2}\nabla\cdot\left(\hat{\kappa}\nabla\theta\right),\text{ and}\label{eq:enthalpy}\\
\nabla\cdot\boldsymbol{u} & = & 0,\label{eq:continuity}
\end{eqnarray}
\end{widetext}
where $Pr=\nu/\kappa_{l}$
is the Prandtl number, {$Ro_c$} is the Rossby number {(see Eq. \ref{eq:Ro_convective})},
and $\hat{\kappa}=\kappa/\kappa_{l}$ is the ratio of the local thermal
diffusivity to the diffusivity in the liquid. The Stefan condition
(Eq. \ref{eq:meltvel}) in nondimensionalized form is given by 
\begin{equation}
u_{m}=\frac{St}{Re\cdot Pr}\left[\hat{\kappa}_{s}\left(\nabla\theta\right)_{s}-\left(\nabla\theta\right)_{l}\right],\label{eq:nondim_meltvel}
\end{equation}
where $\hat{\kappa}_{s}=\kappa_{s}/\kappa_{l}$ is the nondimensional
thermal diffusivity in the solid. Finally, in the
solid there is only heat conduction and hence $\mathbf{u}=0$ in Eqs.
(\ref{eq:momentum}-\ref{eq:enthalpy}). \\

{As the solid melts and the height of the liquid layer increases, the effective Rayleigh and Ekman numbers evolve according to}
\begin{align}
Ra_{\text{eff}} & = Ra\left[\frac{h\left(t\right)}{H/2}\right]^{3},\qquad\textrm{and}\label{eq:Ra_defn}\\
E_{\text{eff}} & = E\left[\frac{H/2}{h\left(t\right)}\right]^{2},\label{eq:Ekman_defn}
\end{align}
respectively, showing that as the solid melts and
the liquid layer becomes deeper, $Ra_{\tiny \text{eff}}$ and $E^{-1}_{\tiny \text{eff}}$ both increase.
We also note that the ratio $\left(Ra/Ra_{c}^{\text{bulk}}\right)_{\text{eff}}\sim RaE^{4/3}$
(from Eq. \ref{eq:Ra_cr_bulk}) increases with time as $h^{1/3}$.

Unless specifically mentioned, we label the results presented here with the reference values $Ra$ and
$E$. The effective
Rayleigh and Ekman numbers $Ra_{\text{eff}}$ and $E_{\text{eff}}$
are considered in the heat transport calculations in \S \ref{subsec:melting_vs_Nusselt}.

Lastly, {the Rossby number $Ro_c$ in Eq. \ref{eq:momentum}, also sometimes called} the convective Rossby number, is a measure of the rotation-dominance of the flow, and is given by
\begin{equation}
Ro_c = \left(\frac{Ra}{Pr Ta}\right)^{1/2}= E\left(\frac{Ra}{Pr}\right)^{1/2},\label{eq:Ro_convective}
\end{equation}
where $Ta=E^{-2}$ is the Taylor number. {Despite system specific definitions of the Rossby number, such as in geophysical fluid dynamics \citep[see e.g.,][Chapter 9]{CushmanBeckers}, all flows with  Rossby numbers much less than unity are rotationally dominated.}

\subsection{Initial and Boundary conditions \label{subsec:ICs_BCs}}

At $t=0$, both the solid and liquid phases are at the melting temperature
$\theta=0$. Unless otherwise mentioned, we use $h_{0}=H/2$. The
upper and lower boundaries are held at temperatures $\theta=-f$ and
$\theta=1$ respectively ($f=0$ except in \S\ref{subsec:vary_kappas}).
The lateral boundaries are insulating, no-slip walls. No-slip conditions
are also applied at the freely evolving phase boundary, where the
temperature is $\theta=0$. {\cite{Ravichandran2020_RotCon} showed
that free-slip boundaries support flow structures that no-slip boundaries cannot.
Here, in order to examine how such structures influence melting dynamics, a}
free-slip velocity condition is used on the lower boundary, except in \S \ref{subsec:noslipbot}, where
we study the influence of the no-slip velocity boundary condition on
the lower boundary.

\subsection{Numerical Simulations \label{sec:Numerical-Simulations}}

Equations (\ref{eq:momentum} - \ref{eq:enthalpy}), together with
Eq. (\ref{eq:nondim_meltvel}), are solved using the finite volume
solver Megha-5 on a uniform grid in all three space directions \citep{prasanth2014,diwan2014,ravichandran2020mammatus,Ravichandran2020_RotCon}.
After every timestep of Eqs. \ref{eq:momentum} and
\ref{eq:enthalpy}, an equilibration step is implemented using Eqs.
\ref{eq:chi} and \ref{eq:EOS_rev}. This procedure is similar to
that used by \cite{RabbanipourEsfahani2018} and has been validated
against analytical results (Appendix A). The requisite velocity conditions
in the resulting arbitrarily shaped solid region are applied using
the volume-penalization method of \cite{kevlahan2001computation},
wherein the solid is modeled as a porous medium with vanishing porosity.
This amounts to adding a term $-\frac{\chi}{\eta}\mathbf{u}$ to the
right hand side of Eq. (\ref{eq:momentum}), where $\eta\ll1$ is
the penalization parameter. Our simulations are performed with 
up to $512^{2}\times256$ gridpoints, a penalization parameter
of {$\eta=2\times10^{-3}$}, and a timestep of {$\delta t = 10^{-3}$}. The
results presented are independent of the grid resolution and insensitive
to the value of the penalisation parameter used (Appendix
B).

We note that for the single component two-phase system considered
here, the solid-liquid interface has to be sharp and hence 
$\chi$ varies smoothly from $0$ to $1$ over a finite number of gridpoints (see
Fig. \ref{fig:chi_theta_vs_z} in Appendix A). For the purposes of
plotting, the solid-liquid interface is taken to be the iso-surface
$\chi=0.5$.

\section{Results and Discussion \label{sec:Results}}

The range of Ekman and Rayleigh numbers we consider
here are listed in Table \ref{tab:parameters}, and correspond to
rapidly rotating convection. For the associated values of $Ra/Ra_{c}$,
we obtain no flow for $Ra<Ra_{c}^{\text{wall}}$; a streaming flow
close to the walls (the `wall modes') for $Ra_{c}^{\text{wall}}<Ra<Ra_{c}^{\text{bulk}}$;
and columnar vortices for $Ra>Ra_{c}^{\text{bulk}}$.
We do not study the geostrophic turbulence regime, $Ra\gg Ra_{c}^{\text{bulk}}$.
In the majority of cases we report here, the flow takes the form of
columnar vortices, with a peripheral retrograde near-wall current.
We show how the nature of the flow controls the morphology
of the melting of the solid, and how the melting influences the flow structures.
We also study the sensitivity of these results to the Stefan number.  As we 
explain below, choosing a Prandtl number of $5$ allows columnar vortices to
form at lower Rayleigh numbers.
\begin{widetext}

\begin{table}
\noindent \begin{raggedright}
(a) 
\par\end{raggedright}
\noindent \begin{centering}
\begin{tabular}{|c|c|}
\hline 
Parameter & Range\tabularnewline
\hline 
\hline 
$E$ & $10^{-3}-8\times10^{-5}$\tabularnewline
\hline 
$Ra$ & $10^{5}-5\times10^{7}$\tabularnewline
\hline 
$Pr$ & $1,5$\tabularnewline
\hline 
$St$ & $0.05-1$\tabularnewline
\hline 
$\tilde{R}=Ra/Ra_{c}^{\text{bulk}}$ & $\mathcal{O}\left(10^{0}\right)-\mathcal{O}\left(10^{2}\right)$\tabularnewline
\hline 
\end{tabular}
\par\end{centering}
\noindent \begin{raggedright}
(b)
\par\end{raggedright}
\noindent \begin{centering}
\begin{tabular}{|c|c|c|}
\hline 
Parameter / boundary & Standard value / type & Special cases\tabularnewline
\hline 
\hline 
Lateral boundaries & Solid walls & Periodic (\S\ref{subsec:Periodic})\tabularnewline
\hline 
Lower boundary & Free-slip & No-slip (\S\ref{subsec:noslipbot})\tabularnewline
\hline 
$\hat{\kappa}_{s}$ & $1$ & $0.2,5$ (\S\ref{subsec:vary_kappas})\tabularnewline
\hline 
\end{tabular}
\par\end{centering}
\caption{\label{tab:parameters} (a) Ranges of the controlling
parameters (defined in the text) used. (b) Typical boundary
conditions or values of parameters used, except in special
cases called out in the text.}
\end{table}
\end{widetext}

\subsection{Flow structure and melting morphology \label{subsec:Flow-structures}}

Before discussing the influence of these flow structures on the morphology
of the melting, we look first at some properties of the columnar vortices
themselves. We identify these vortices as isolated
regions at the horizontal plane given by $z=H/4$ where 
\[
\omega_{z}=\frac{\partial v}{\partial x}-\frac{\partial u}{\partial y}> \omega_{0}.
\]
Whilst the threshold used, $\omega_{0}=0.25$, is arbitrary, this choice does not change the number of vortices significantly,
but it does affect the vortex area, as is to be expected. The rotating convection driving the melting is time dependent, and the mean and maximum vorticity increase with time. For this reason, we rationalize an arbitrary threshold in order to have a means of comparing
vortex areas and numbers at different points of time.

\subsubsection{Rotational Dominance and columnar vortices}

\begin{figure}
\noindent \centering{}\includegraphics[width=1\columnwidth]{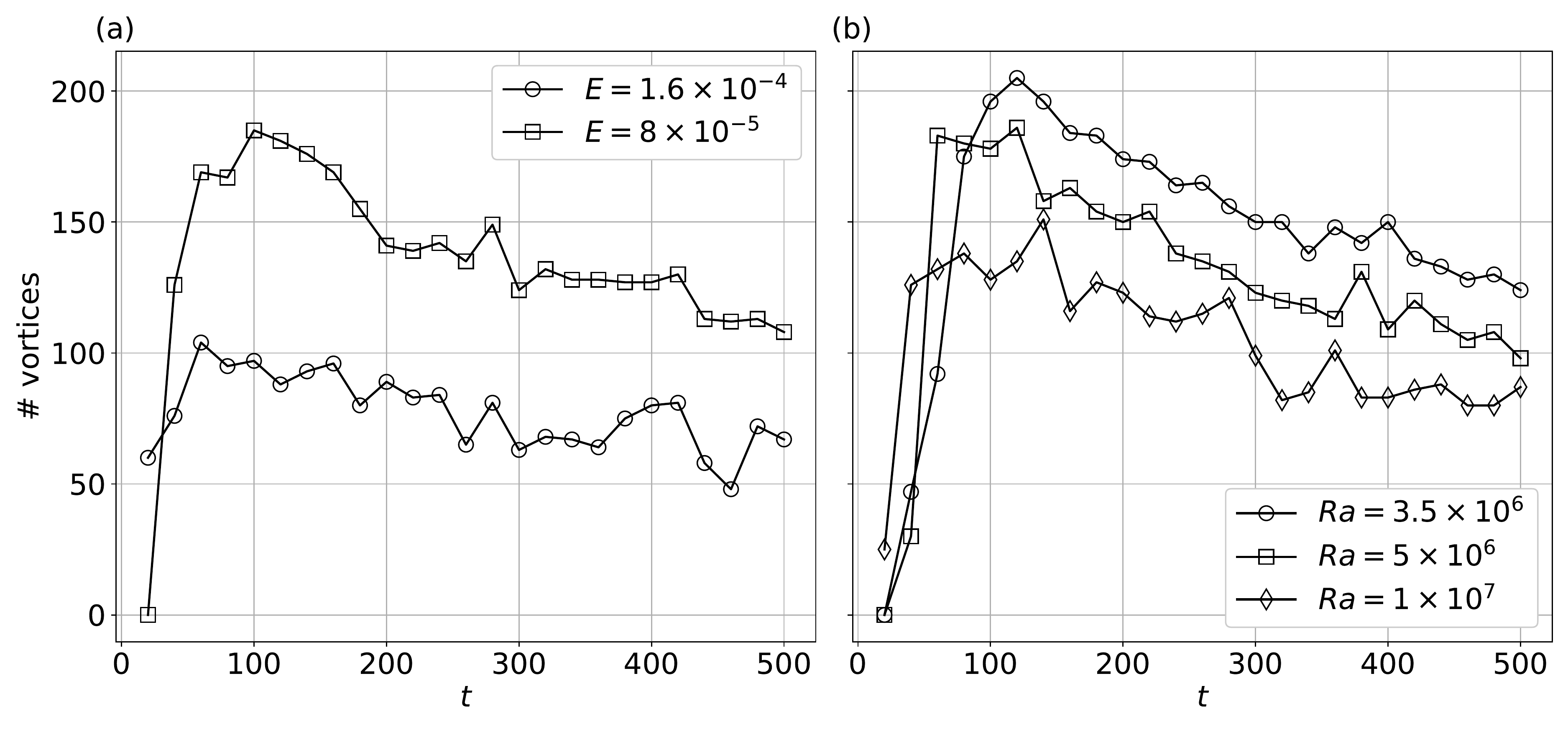}
\caption{\label{fig:nvors_vs_t} The number of columnar vortices as a function of time for $Pr=5$, $St=1$ and (a) $Ra=7.8\times10^{6}$, and (b) $E=10^{-4}$, showing that as rotational effects become 
more dominant the number of vortices increases.}
\end{figure}

\begin{figure}
\includegraphics[width=1\columnwidth]{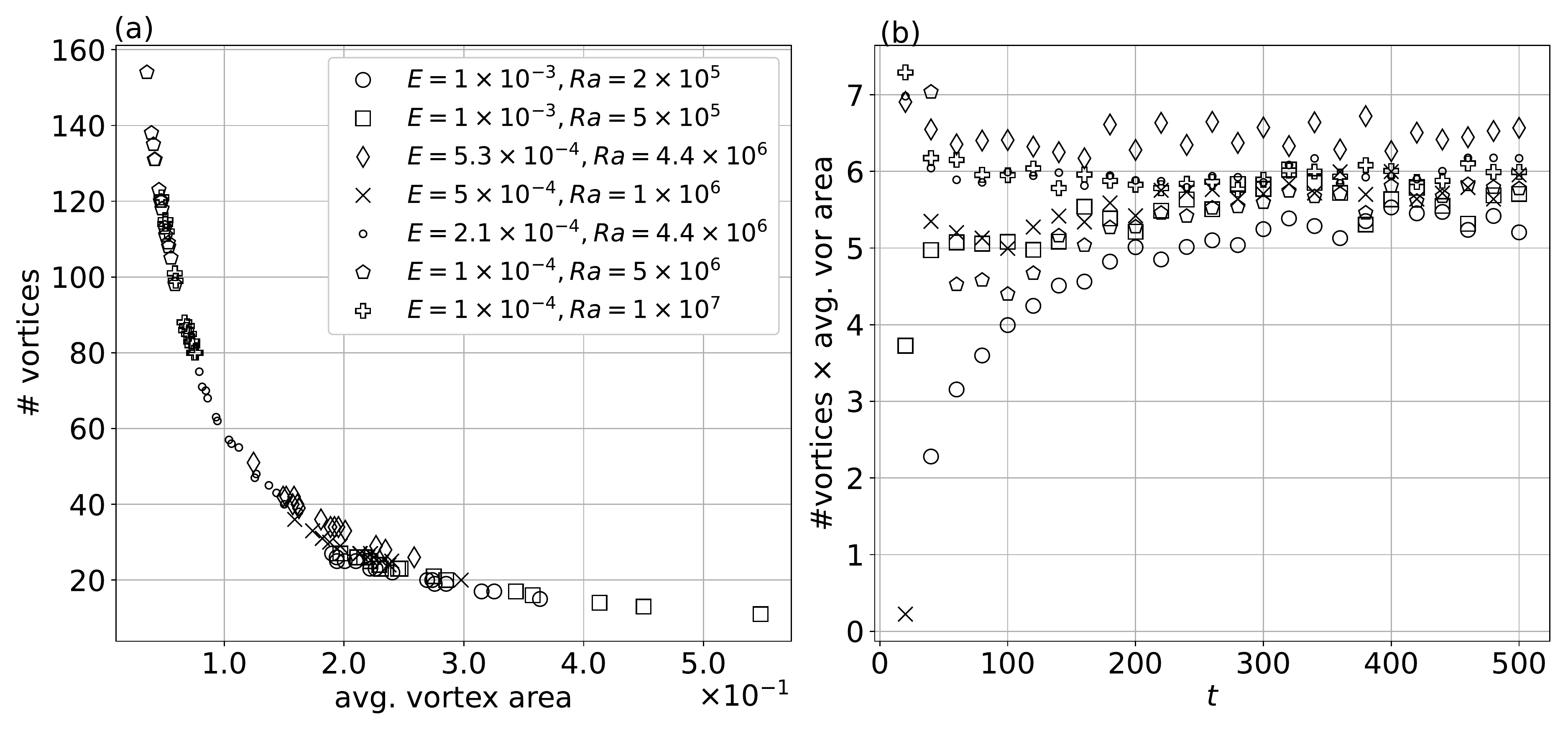} 
\caption{\label{fig:nvor_vs_vorarea} Dependence of flow structure on the flow
parameters {$E$ and $Ra$ with $Pr=5$ and $St=1$}. (a) The number of vortices and the average area of each
vortex area inversely proportional to each other. (b) The total cross-sectional
area of the columnar vortices is an increasing function of time before
saturating at late times. }
\end{figure}

For a given $Ra, Pr$ combination, decreasing
$E$ increases the rotational control of the flow
and we expect a larger
number of thinner vortices \citep{zhong1991asymmetric,Sakai1997,VorobieffEcke1998}, as shown in Fig. \ref{fig:nvors_vs_t}(a).
Moreover, as the Rayleigh number increases the number of vortices decreases, as shown in Fig. \ref{fig:nvors_vs_t}(b).
Of particular relevance to the phase-change dynamics, Fig. \ref{fig:nvor_vs_vorarea}(a)
shows that as the number of vortices increases the average area of
each vortex decreases. Moreover, this behavior is independent of the
flow regimes studied, as evidenced by the parametric collapse onto
a single curve. Figure \ref{fig:nvor_vs_vorarea}(b) shows that, beyond
the initial transients, the total vortex area reaches a quasi-steady state. For a given $E$, this total
vortex area increases with increasing $Ra$.

\begin{figure}
\includegraphics[width=1\columnwidth]{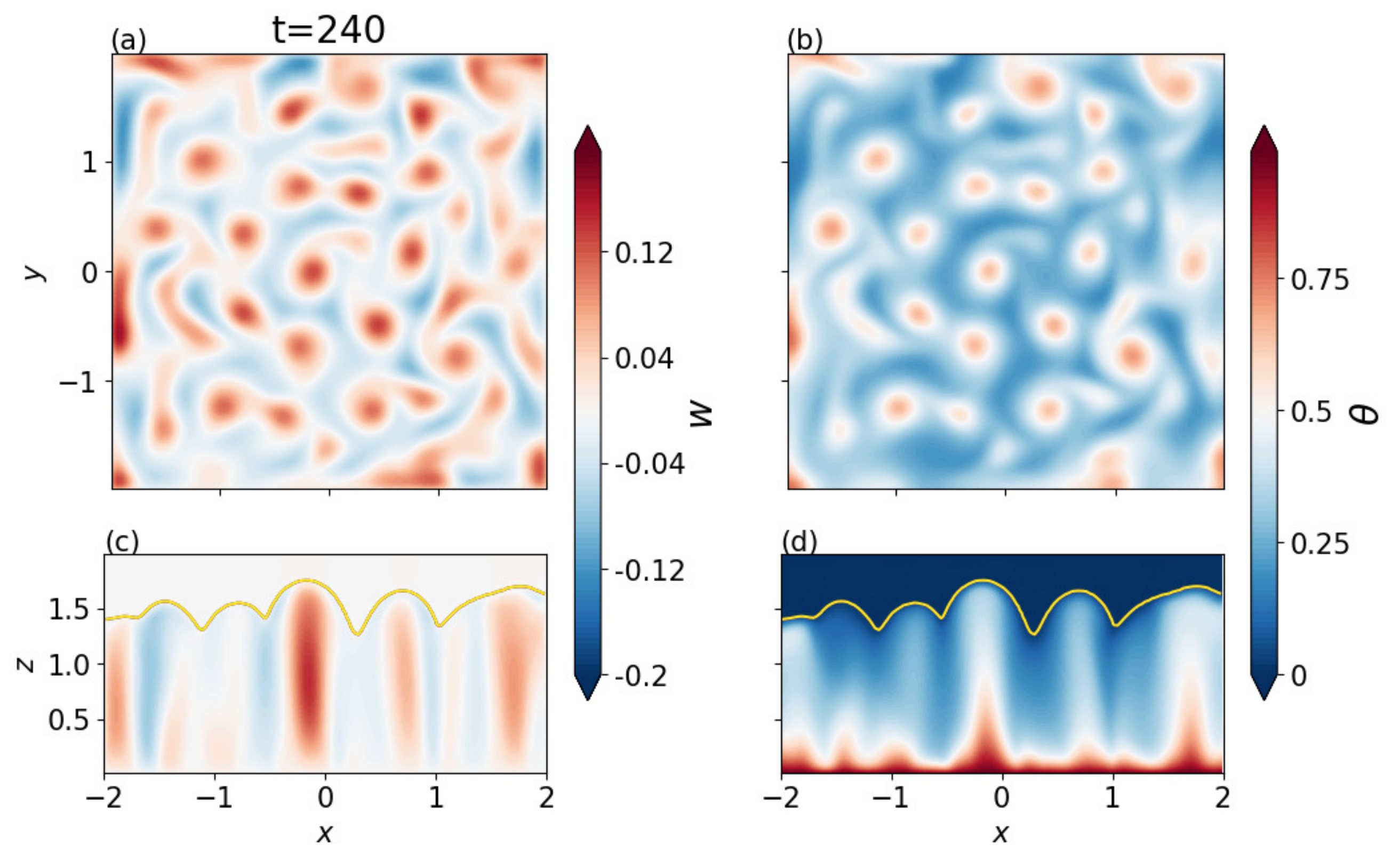}

\includegraphics[width=1\columnwidth]{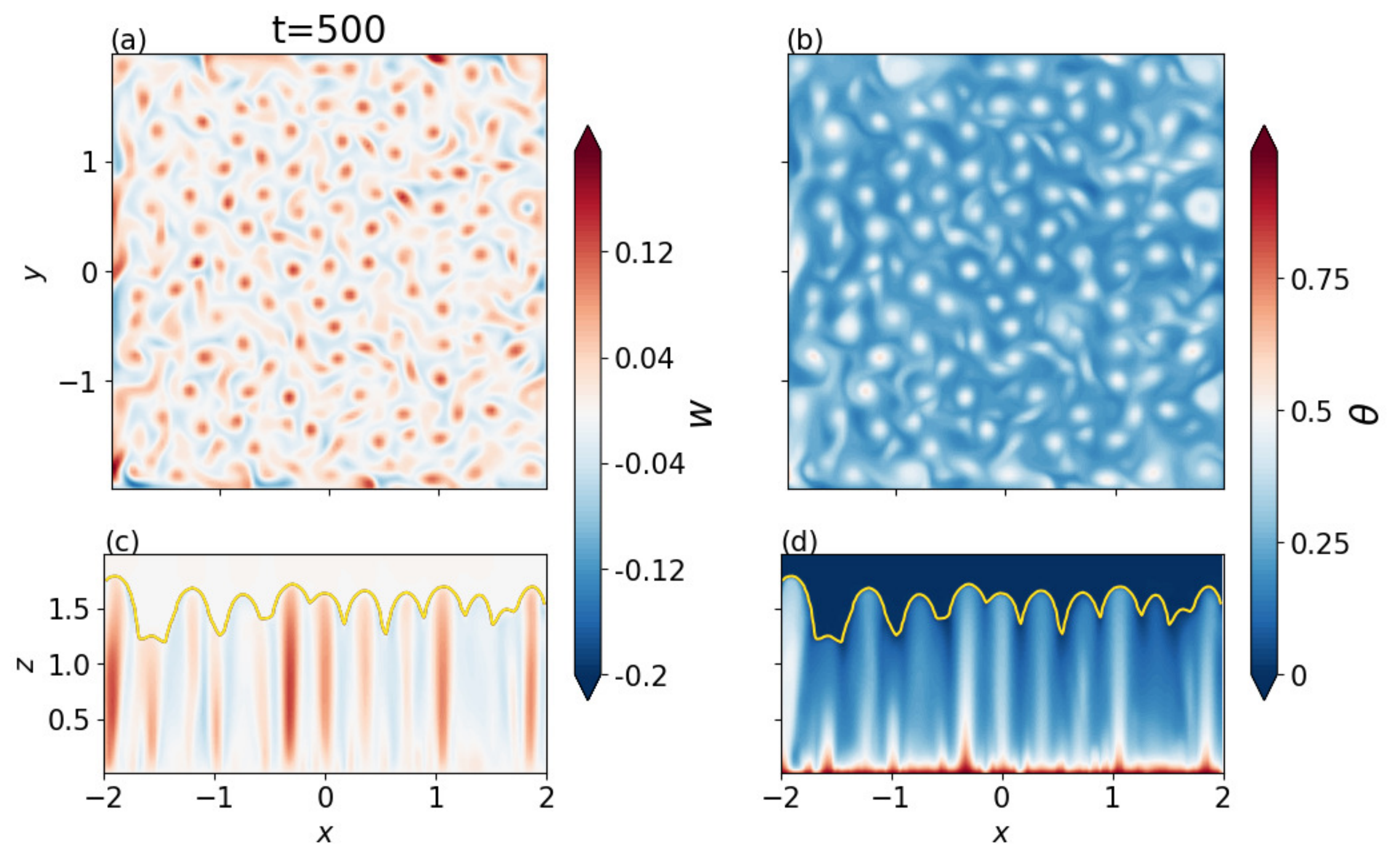}

\caption{\label{fig:cross_sections_vs_E_Ra} Cross sections of the temperature
$\theta$ and the vertical velocity $w$ for (top four panels) {$E=10^{-3}$, $Ra=2\times10^{5}$},
$Pr=5$, {$f=0$}, {$St=1$, $t=240$}; and (bottom four panels) {$E=8\times10^{-5}$, $Ra=7.8\times10^{6}$,}
$Pr=5$, {$f=0$}, {$St=1$, $t=500$}. In each subfigure, the horizontal sections (a,b)
are plotted on the $z=H/4$ plane and the vertical sections (c,d)
are plotted on the $y=0$ plane. The yellow lines in the vertical
sections show the instantaneous location of the solid-liquid interface.
Vertical heat transport occurs in columnar vortices as reflected in
the pattern of the melting solid.}
\end{figure}

Vertical and horizontal cross-sections of the temperature and vertical
velocity in Fig. \ref{fig:cross_sections_vs_E_Ra} show the typical
patterns of flow and melting seen at the smallest and largest $E$
in our simulations (Table \ref{tab:parameters}).  Particularly notable is the increase
in the number of vortices in smaller $E$ more rotationally dominant 
flows.

\begin{figure}
\includegraphics[width=1\columnwidth]{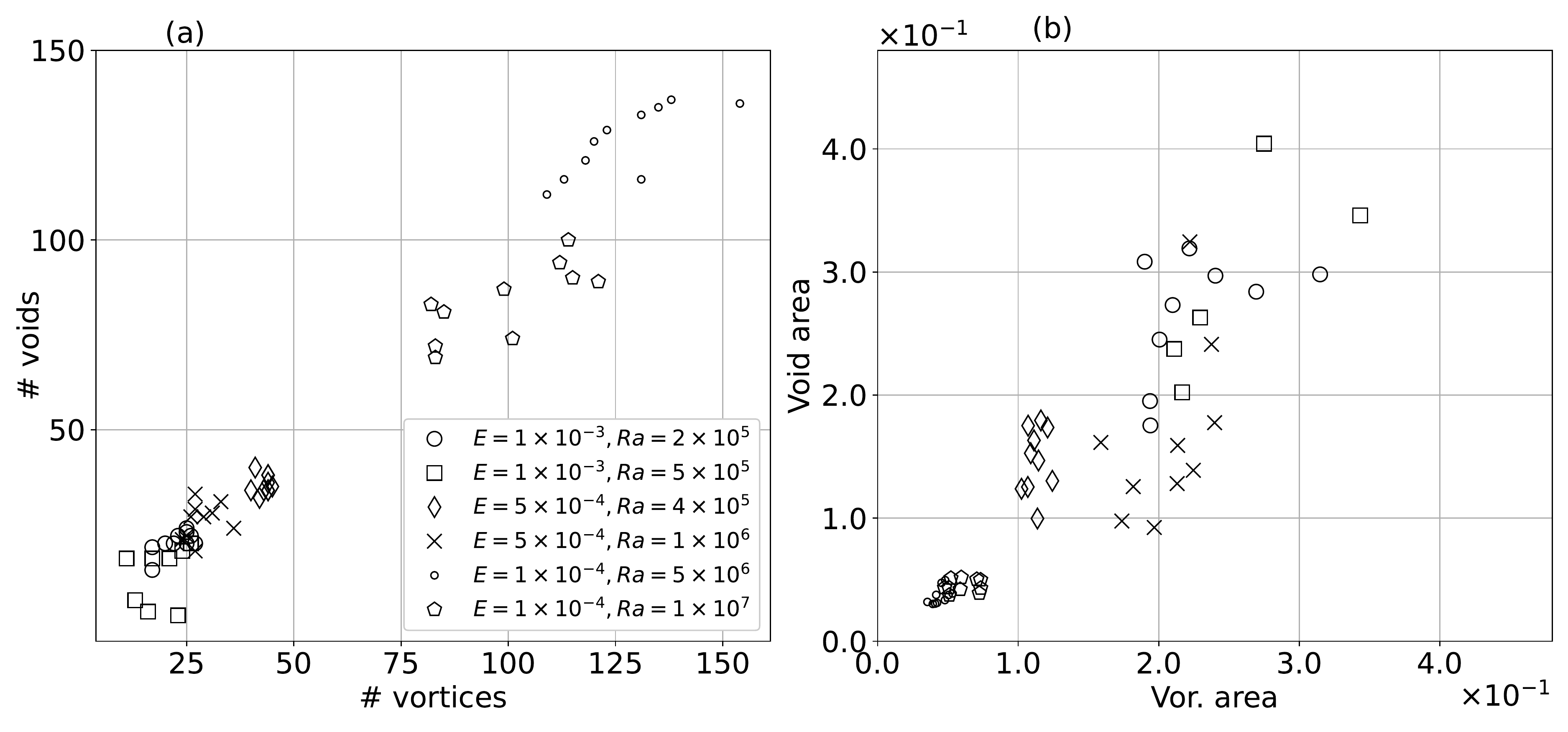}
\caption{\label{fig:nvors_vs_nholes} (a) The number of solid voids as a function
of the number of vortices, showing the linear dependence of the former
on the latter. (b) The area of the solid voids as a function of the
area of the vortices. In both figures, points are plotted every $10$
flow units excluding initial transients and before the fluid comes
into direct contact with the upper boundary. {$Pr=5$, {$f=0$}, and $St=1$ in all cases shown.}}
\end{figure}

These columnar vortices carry heat from the lower boundary to the
solid and, as Figs. \ref{fig:cross_sections_vs_E_Ra} show, etch voids
into the solid. Therefore, the morphology of the
phase boundary---the average area and number of void regions melted
into the solid---reflects the state of the flow.
Fig. \ref{fig:nvors_vs_nholes} shows that the number of voids and their
average cross-sectional area are proportional to the number and the
average area of the vortices respectively. However, whereas the number
and size of the vortices play a role in the total heat transport by
the fluid, the heat transfer is not simply proportional to the total
vortex area, but depends additionally upon their specific heat and
velocity, as described presently.

{We note that Fig. \ref{fig:cross_sections_vs_E_Ra} shows sharp cusps in the solid-liquid interface. Such cusps are a common challenge in 
numerical simulations of interfacial flows \citep[e.g.,][]{Popinet2018ARFM}.  Here, we find no evidence that these features influence the
overall dynamics appreciably.  In particular, we have verified that the shapes and sizes of the cusps, and the shapes
and areas of the voids are independent of grid resolution.}

\begin{figure}
\noindent \centering{}\includegraphics[width=1\columnwidth]{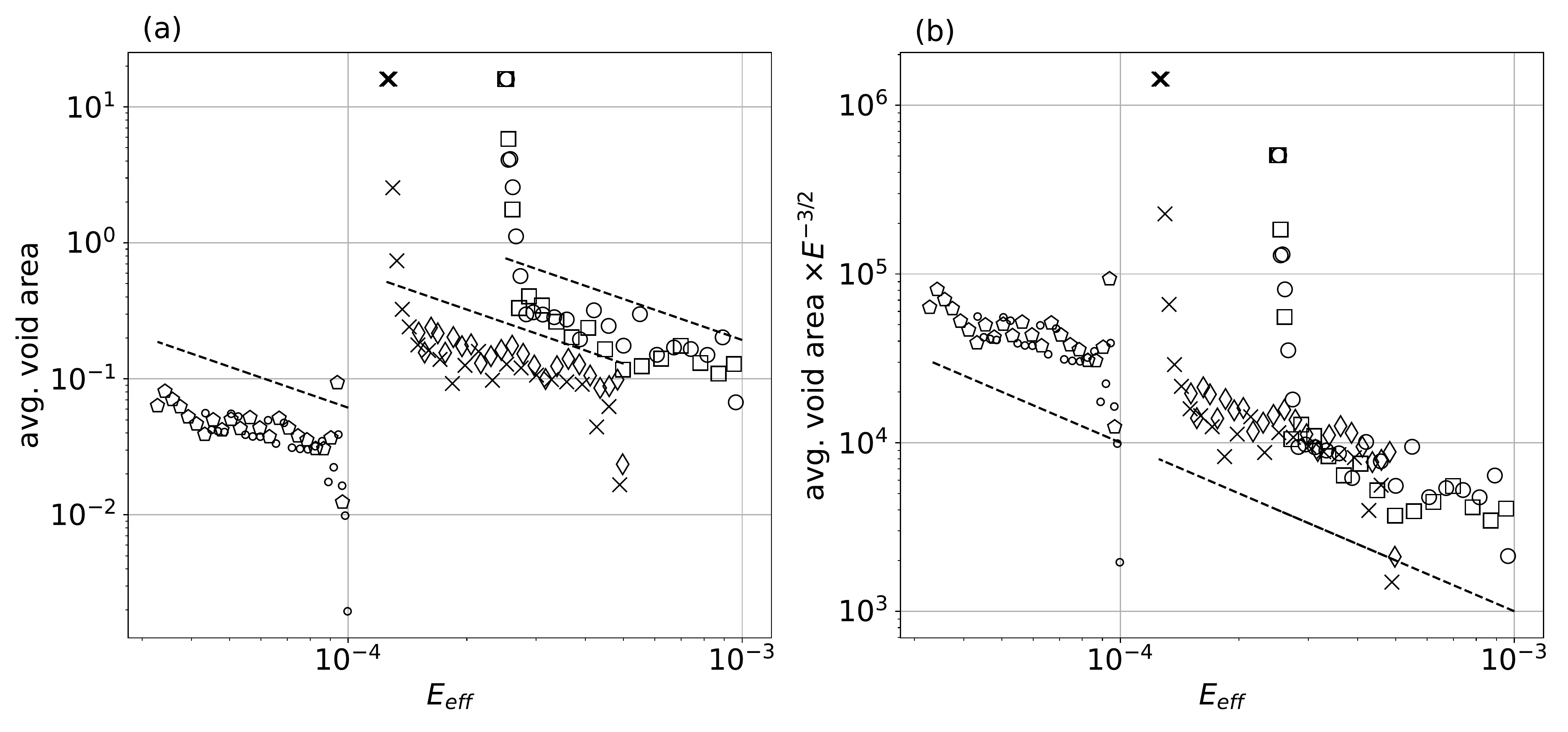}
\caption{{\label{fig:holearea_vs_Eeff} (a) The average area of the voids formed
grows with time, and is seen to grow proportionally to $E_{\tiny \text{eff}}$.
Apart from the initial transients (and the divergence to infinity
in cases where all the solid has melted away within the simulation
time), the same proportionality holds for different values of $E={E_{\tiny \text{eff}}(t=0)}$.
(b) A resonable collapse is obtained if the void areas are multiplied
by $E^{-3/2}$ (note that we multiply by the \emph{initial} value,
not the abscissa). The parameter combinations are the same as in Fig.
\ref{fig:nvors_vs_nholes}, and the symbols have the same meaning.}  }
\end{figure}

{As the melting proceeds and the height of the liquid layer grows,
$Ra_{\text{eff}}$ and $E_{\text{eff}}^{-1}$ grow as well (Eqs. \ref{eq:Ra_defn}
and \ref{eq:Ekman_defn}). Moreover, as the vortices merge into larger vortices, the voids do as well.
The average area of the voids thus grows as a function
of time, as seen in the plot of the average void area versus $E_{\text{eff}}$
in Fig. \ref{fig:holearea_vs_Eeff}. 
We note, however, that Fig. \ref{fig:holearea_vs_Eeff} is primarily intended to motivate future work.  Namely, because 
they do not span two decades on both axes, a rigorous evaluation \citep[see, e.g.,][]{Stumpf2012} of the relationship between the void area and $E_{\tiny\text{eff}}$
cannot be made.}

\begin{figure}
\includegraphics[width=0.48\columnwidth]{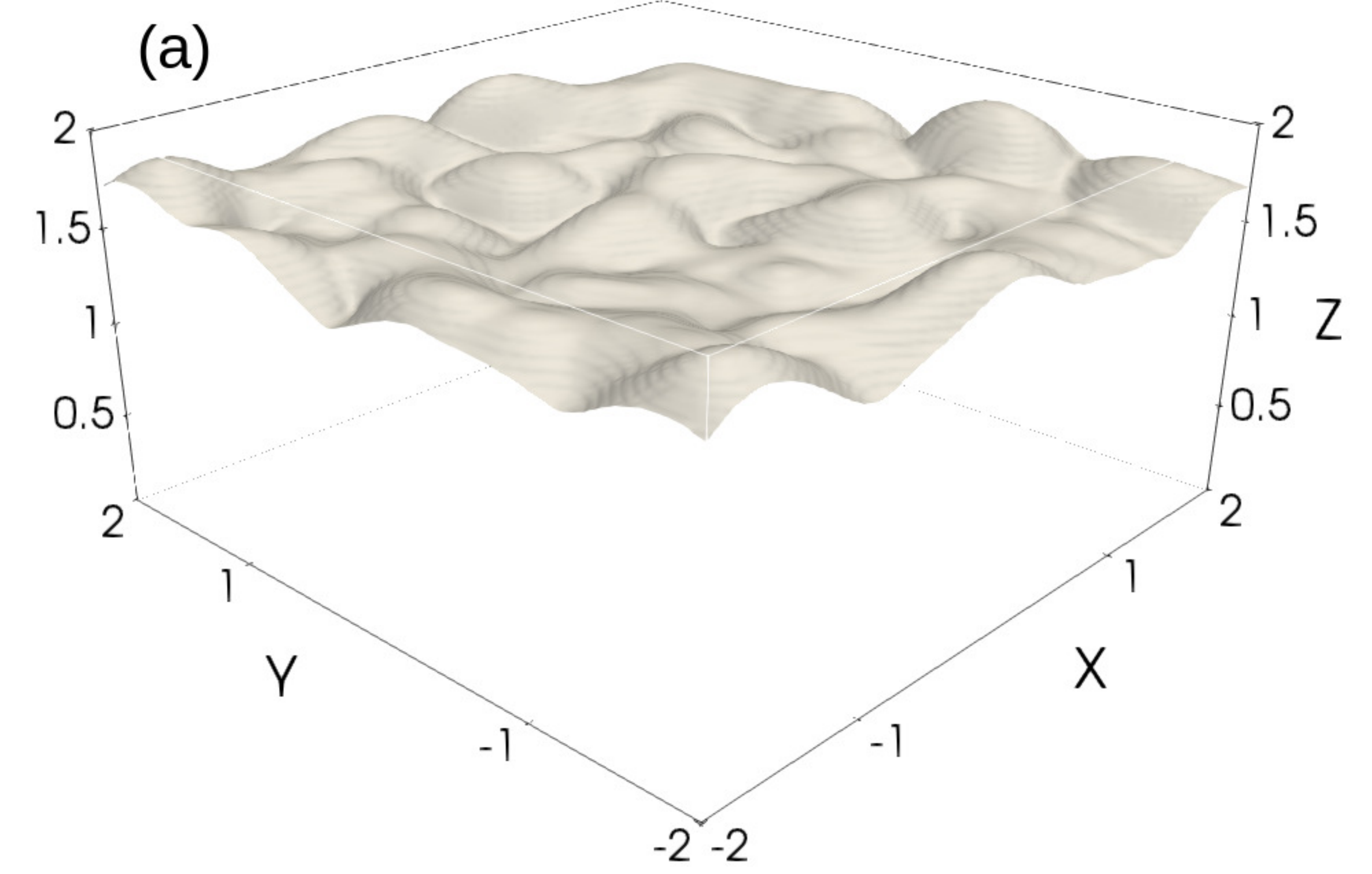}
\includegraphics[width=0.48\columnwidth]{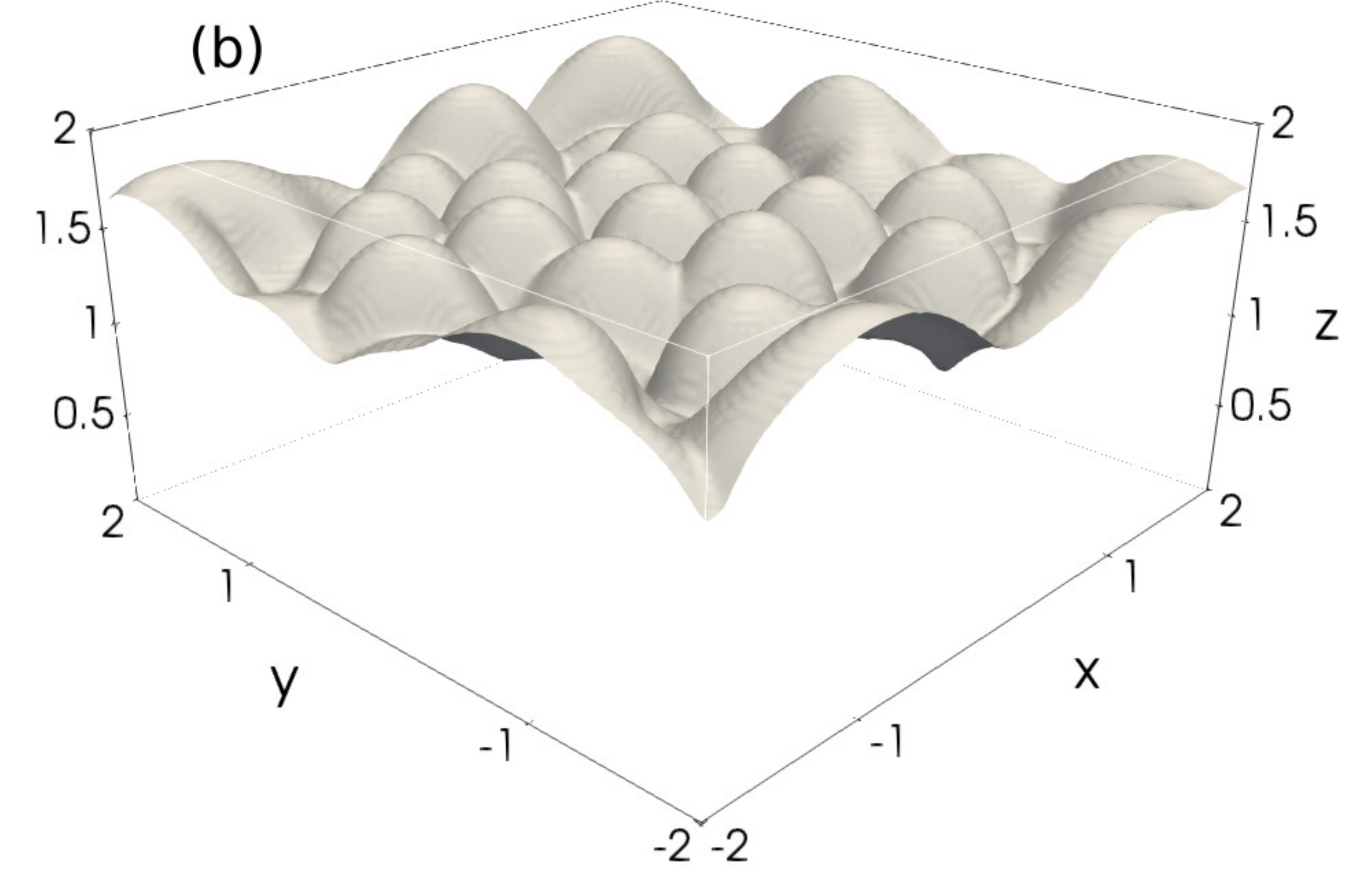}

\includegraphics[width=0.48\columnwidth]{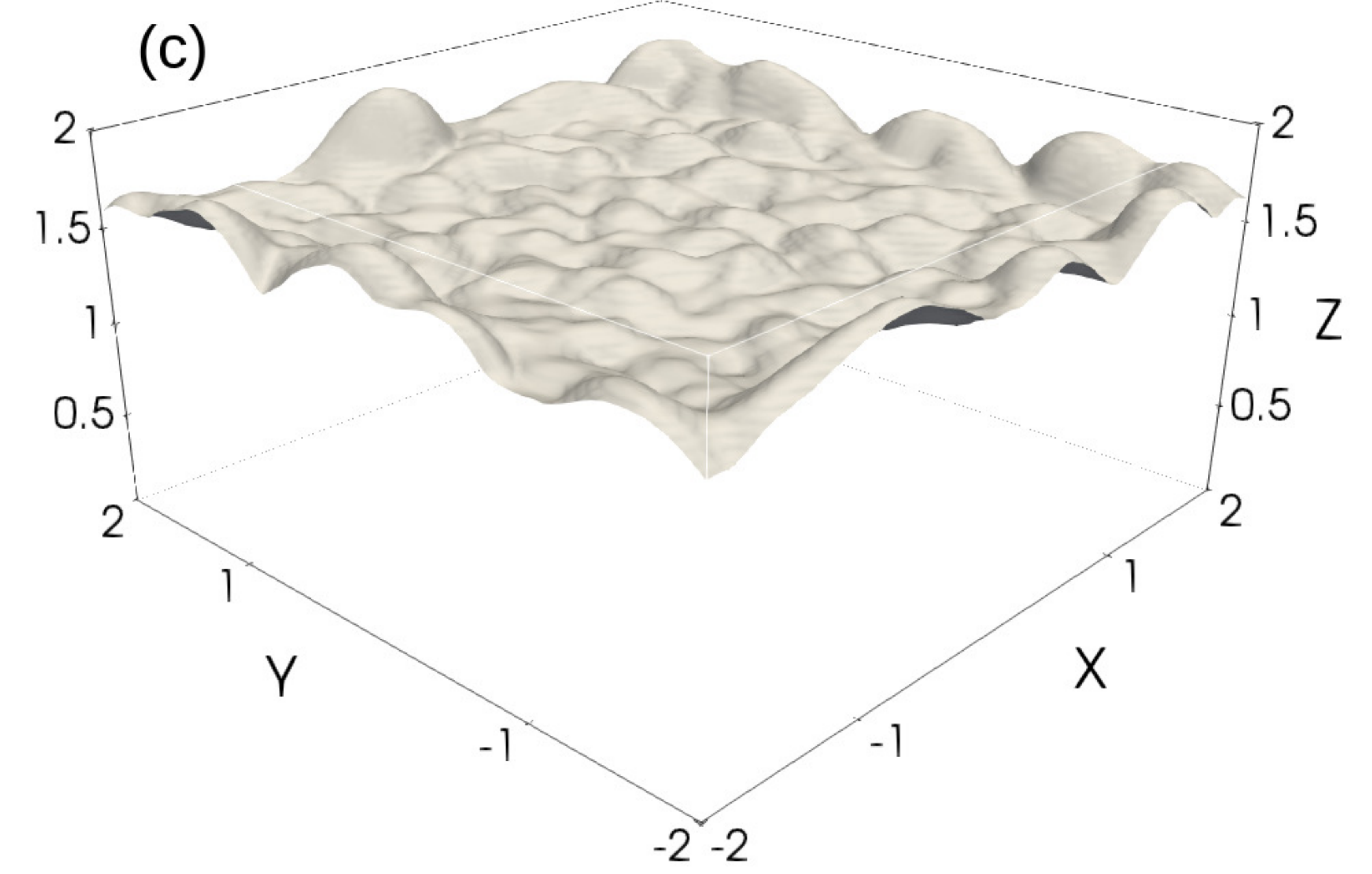}
\includegraphics[width=0.48\columnwidth]{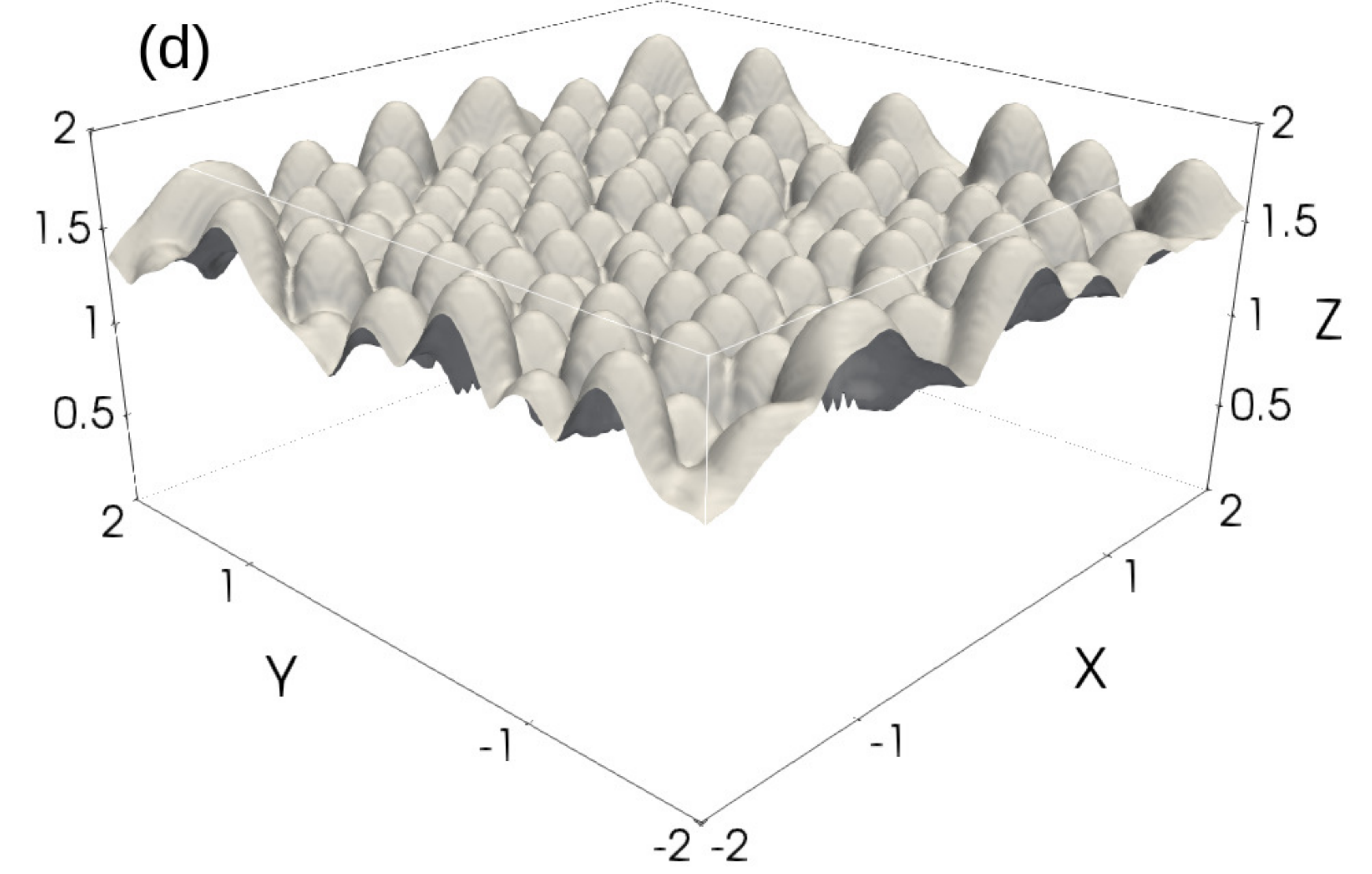}

\caption{\label{fig:interface_vs_Pr}The solid-liquid interface (viewed from
the solid side) {with $St=1$, $f=0$} for (a) {$E=10^{-3}$, $Ra=2\times10^{5}$}, $Pr=1$,
{$t=120$}; (b) {$E=10^{-3}$, $Ra=2\times10^{5}$,} $Pr=5$, {$t=240$};
(c) {$E=10^{-4}$, $Ra=5\times10^{6}$,} $Pr=1$; {$t=240$}; (d) {$E=10^{-4}$,
$Ra=5\times10^{6}$,} $Pr=5$, {$t=500$}. For $Pr=5$, vertical heat
transport occurs in columnar vortices as reflected in the pattern
of the melting solid. }
\end{figure}

{The convective Rossby number, $Ro_{c}$, is another key parameter that 
quantifies the rotational control of the flow.  As seen in Eq. \ref{eq:Ro_convective}, 
for a given combination of $Ra$ and $E$, a larger $Pr$ leads to a smaller $Ro_{c}$, and thus to greater
rotational dominance.} In Fig. \ref{fig:interface_vs_Pr}, this is reflected in the melt voids 
that are created by the columnar vortices present for $Pr=5$, but absent for $Pr=1$.

\begin{figure}
\begin{centering}
\includegraphics[width=1\columnwidth]{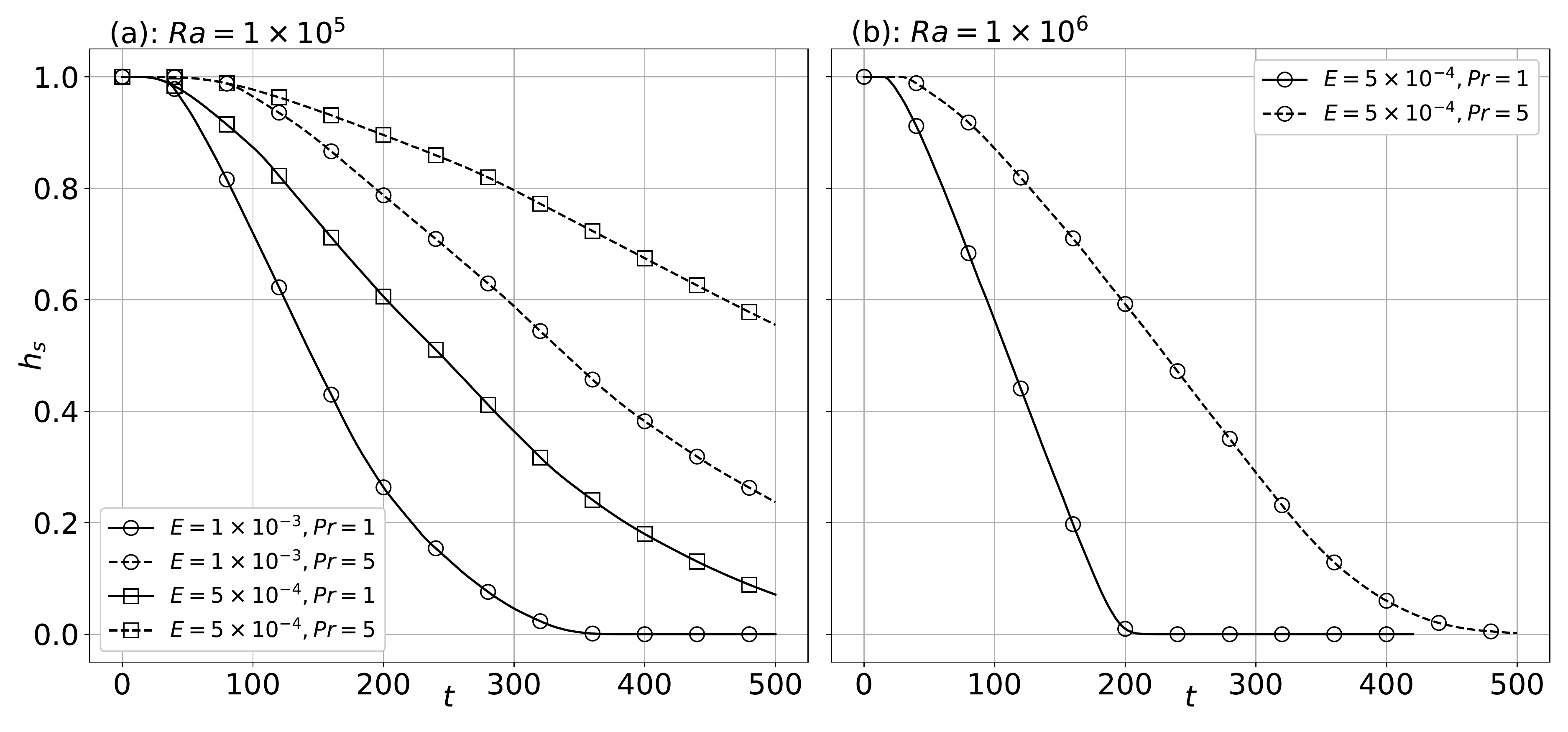}
\par\end{centering}
\caption{\label{fig:chi_vs_t_Pr} The volume averaged height of the solid $h_{s}=H-h$
as a function of time, showing the role of the flow parameters, with
$St=1$, $f=0$. (a) Columnar vortices are absent for both Prandtl
numbers. (b) Columnar vortices are present for $Pr=5$. For the three
combinations of $E,Ra$ (i) $E=10^{-3}$, $Ra=10^{5}$, $Ra/Ra_{c}^{\text{bulk}}=4.2$;
(ii) $E=5\times10^{-4}$, $Ra=10^{5}$, $Ra/Ra_{c}^{\text{bulk}}=1.6$;
(iii) $E=5\times10^{-4}$, $Ra=10^{6}$, $Ra/Ra_{c}^{\text{bulk}}=16.6$.
For a given $Ra$, melting is slower for larger $Pr$ regardless of
the degree of supercriticality $Ra/Ra_{c}$ or the presence of columnar
vortices.}
\end{figure}

{We note that the times at which the phase boundaries are shown {in} Fig. \ref{fig:interface_vs_Pr} reflect that for a given $Ra$, a reduction in $Pr$ 
reflects an increase in heat transfer and hence melt rate, further in evidence of which is seen
in Fig. \ref{fig:chi_vs_t_Pr}, where we plot the amount of solid $h_{s}\left(t\right)=H-h$ as a function
of time for $Pr=1$ and $Pr=5$.}

\begin{figure}
\begin{centering}
\includegraphics[width=1\columnwidth]{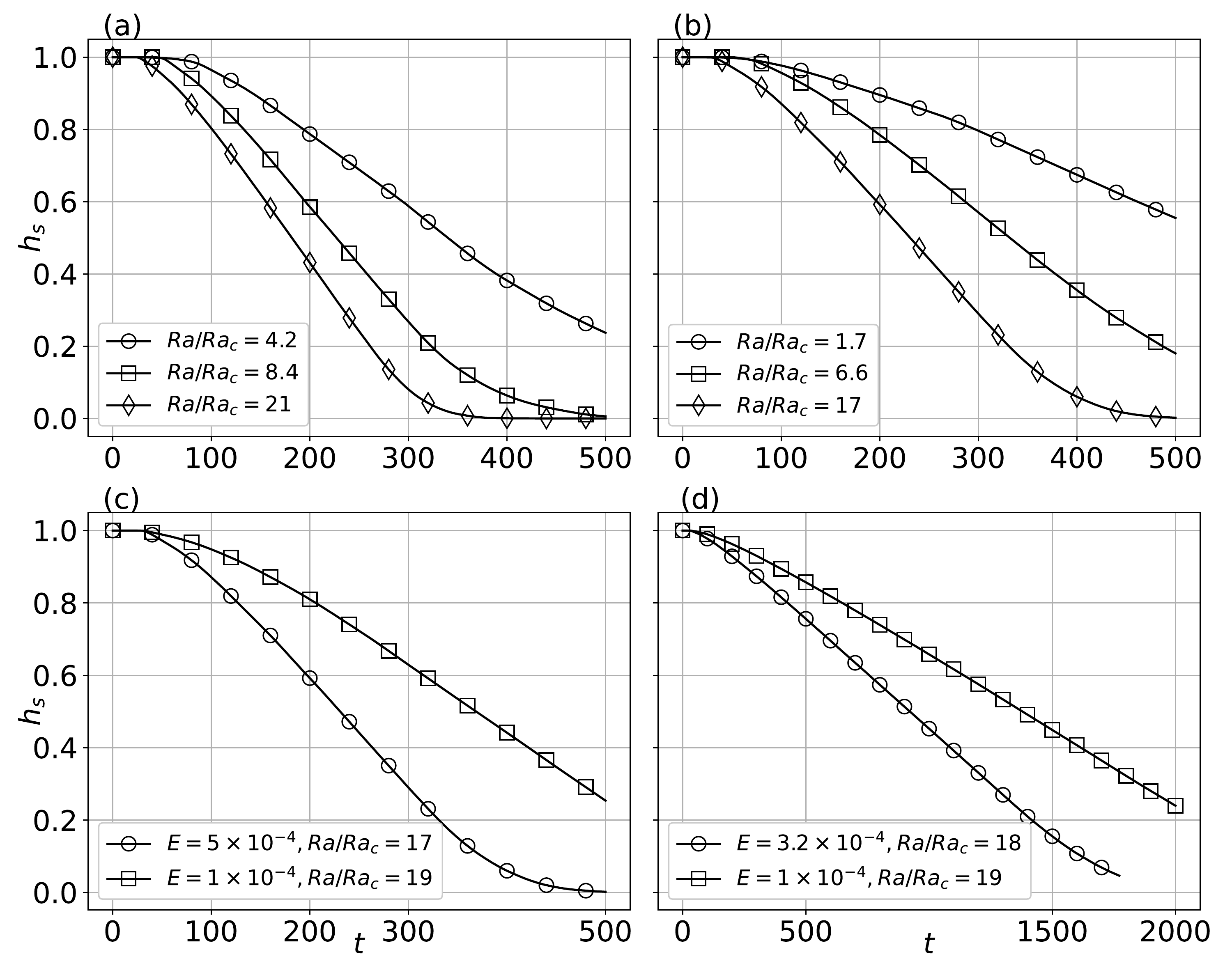}
\par\end{centering}
\caption{\label{fig:chi_vs_t_Ra} The volume averaged height of the solid $h_{s}$
as a function of time, showing {the role of the Ekman and Rayleigh
numbers} for $Pr=5$, {$f=0$}, and $St=1$ (a,b,c) and $St=0.2$ (d). {Increasing
$Ra/Ra_{c}$ leads to a larger melt rate, as shown for (a) $E=10^{-3}$
and (b) $E=5\times10^{-4}$. For comparable $Ra/Ra_{c}$, melting
is slower for smaller $E$, as seen in (c) and (d).} Note that the simulations
in (d) are run for {$2000$} flow time units.}
\end{figure}

It is intuitive that for a given $E$, the melt rate increases with $Ra$ and this is seen in Fig. \ref{fig:chi_vs_t_Ra}(a) and (b).
Moreover, for similar values of $Ra/Ra_{c}$, melting
is faster for larger $E$, when vertical transport is less rotationally
constrained. We analyze the energy balance underlying the melting
rates and the effective Nusselt numbers in detail in \S \ref{subsec:melting_vs_Nusselt}.

\subsubsection{Wall modes and peripheral melting}

\begin{figure}
\includegraphics[width=1\columnwidth]{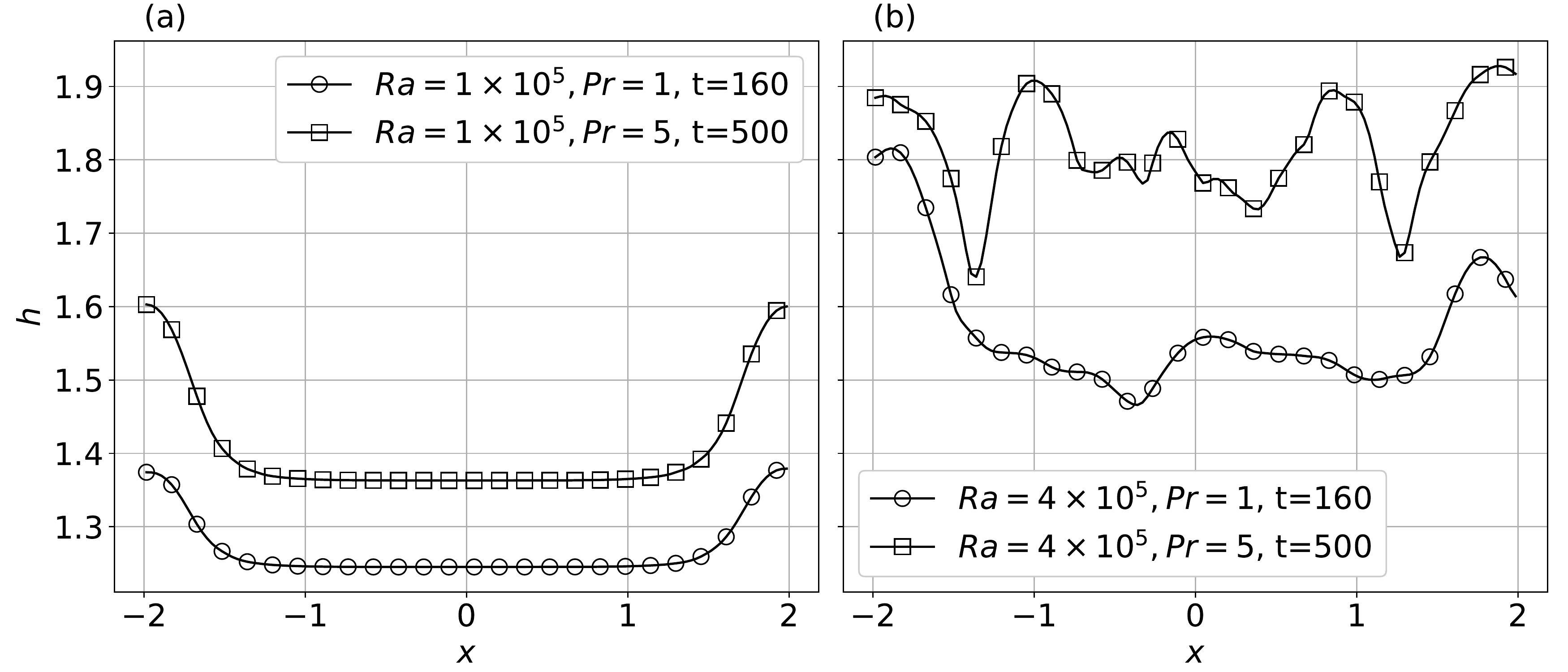}
\caption{\label{fig:liq_height_vs_xy} The height of the
liquid layer, averaged in the horizontal direction {for $y\in[-H/2,H/2]$}, as a function of the horizontal coordinate $x$ for $E=5\times10^{-4}$,
$St=1$ and {$f=0$}. The influence of the peripheral current is larger when the
Rayleigh number is close to the critical Rayleigh number for flow
in the bulk (Eq. \ref{eq:Ra_cr_bulk}). Here we have $Ra_{c}^{\text{bulk}}=6\times10^{4}$,
$Ra_{c}^{\text{wall}}=6.3\times10^{4}$, giving (a) $Ra/Ra_{c}^{\text{bulk}}=1.66$,
$Ra/Ra_{c}^{\text{wall}}=1.57$, and (b) $Ra/Ra_{c}^{\text{bulk}}=6.6$,
$Ra/Ra_{c}^{\text{wall}}=6.3$.}
\end{figure}

When the Rayleigh number approaches the critical value, $Ra_{c}^{\text{bulk}}$,
heat is transported predominantly through the peripheral streaming
current, and hence the solid regions closer to the walls melt significantly faster than the interior, which, as shown in Fig. \ref{fig:liq_height_vs_xy}, remains more planar.
Whereas in Fig.\ref{fig:liq_height_vs_xy}(a), $Ra/Ra_{c}$ = $\mathcal{O}\left(1\right)$, as it increases we see both the effects of the wall modes
and the bulk flow. Thus, when columnar vortices are present, as is the case for $Pr=5$
in Fig. \ref{fig:liq_height_vs_xy}(b), the voids formed penetrate
deeper into the solid than the melt regions created by the wall modes.

\subsubsection{Initial fluid layer height}

\begin{figure}
\includegraphics[width=1\columnwidth]{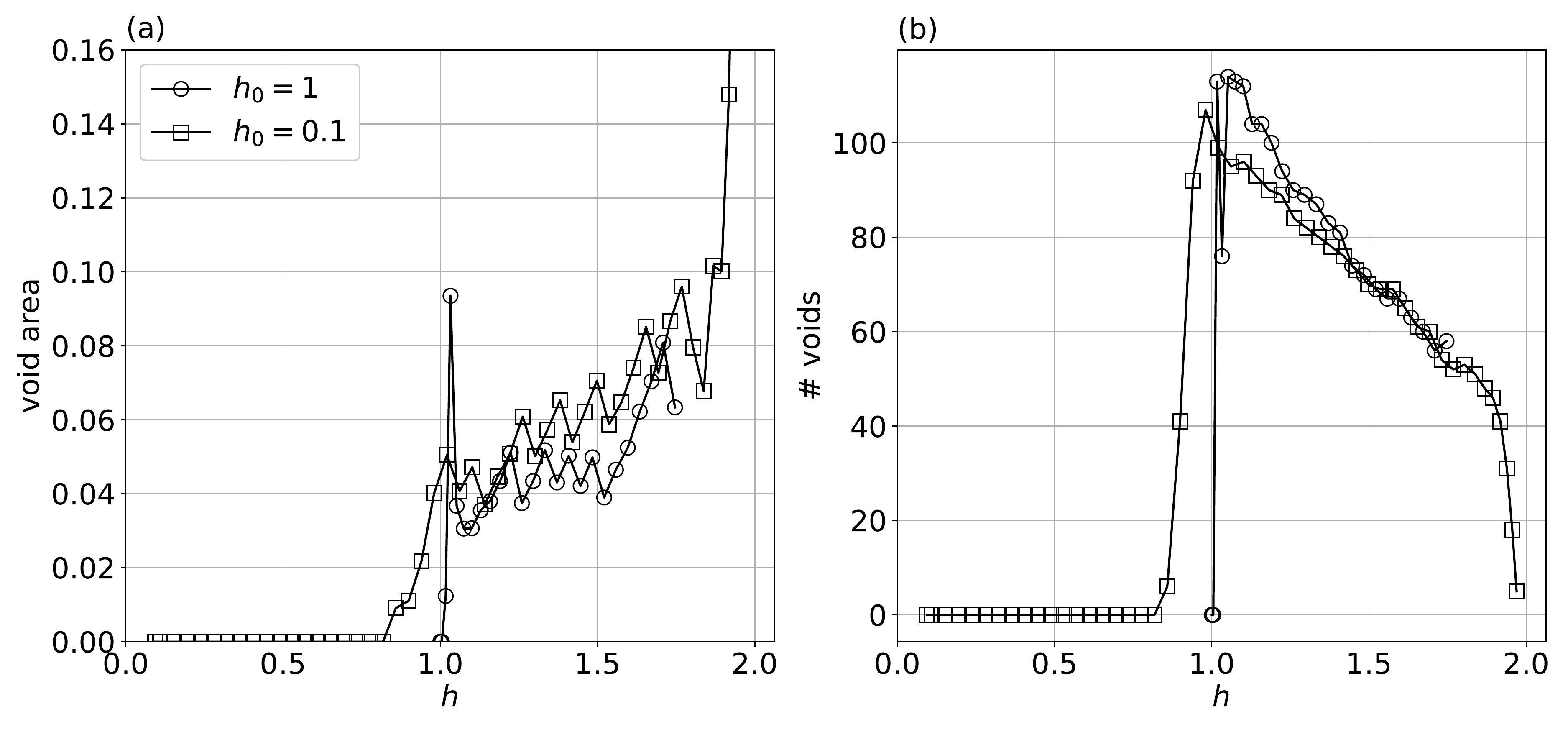}
\caption{\label{fig:nvoids_voidareas_vs_h0} (a) The number of solid voids,
and (b) The area of the solid voids as a function of the liquid height
$h$, for $E=10^{-4}$, $Ra=10^{7}$, $Pr=5$, {$St=1$, $f=0$}. }
\end{figure}

The effective Rayleigh number at $t=0$ is determined by the initial height of the liquid $h_{0}$.  In recent studies of convection-driven melting \citep[e.g.,][]{RabbanipourEsfahani2018,Favier2019}, the initial liquid height is taken to be small fraction of the domain height $H$, such that $Ra_{\text{eff}}(t=0)<Ra_{c}$.
Thus, convection begins only after an initial stage where melting occurs by the relatively slow diffusion of heat, which eventually leads to $Ra_{\text{eff}}>Ra_{c}.$ In our simulations
with $h_{0}=H/2$, the initial $Ra$ is sufficiently large so that convection occurs immediately. 
While the melting history will obviously depend on $h_{0}$, this choice does not change the general conclusions drawn from our simulations.
We show this in Fig. \ref{fig:nvoids_voidareas_vs_h0} by comparing the void area and number as a function of the height of the fluid layer
in simulations with {$h_{0}=1$ and $h_{0}=0.1$}. Apart from initial transient differences, the curves follow very similar trajectories.\\

\subsubsection{Stefan number}

\begin{figure}
\includegraphics[width=0.8\columnwidth]{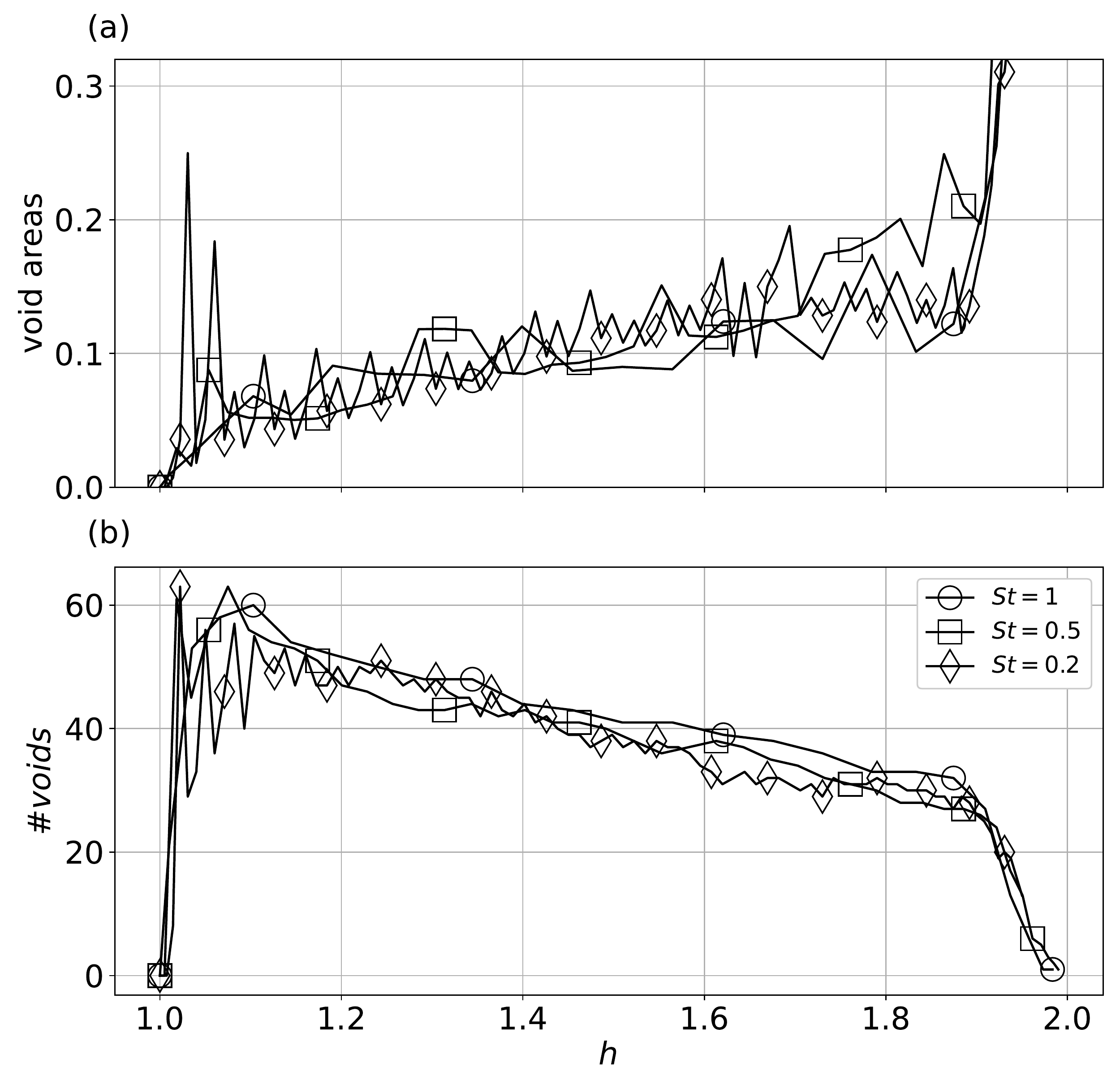}

\caption{{\label{fig:nvoids_voidareas_vs_St}(a) The number of solid voids;
and (b) The area of the solid voids as a function of the liquid height
$h$, for $E=3.2\times10^{-4}$, $Ra=2\times10^{6}$, $Pr=5$, {$f=0$}.}}
\end{figure}

Smaller Stefan numbers, as defined in Eq. \ref{eq:Stefan}, are associated with large latent heats and thus lead to lower melt rates \citep[see e.g., ][] {worster2000solidification}, in which case simulations need to be run for longer times.  However, the melting morphology we find is independent of Stefan number for the range studied ($St$ = 0.2 to 1), which is shown by 
plotting the number and areas of the voids formed in Fig. \ref{fig:nvoids_voidareas_vs_St}. The same is found in the melting of pure solids driven by non-rotating convection \citep{RabbanipourEsfahani2018,Favier2019}.

\subsection{Special cases \label{subsec:Special-cases}}

\subsubsection{No-slip lower boundary: melting rates, flow structures and wall modes \label{subsec:noslipbot}}

\begin{figure}
\noindent \begin{centering}
\includegraphics[width=0.6\columnwidth]{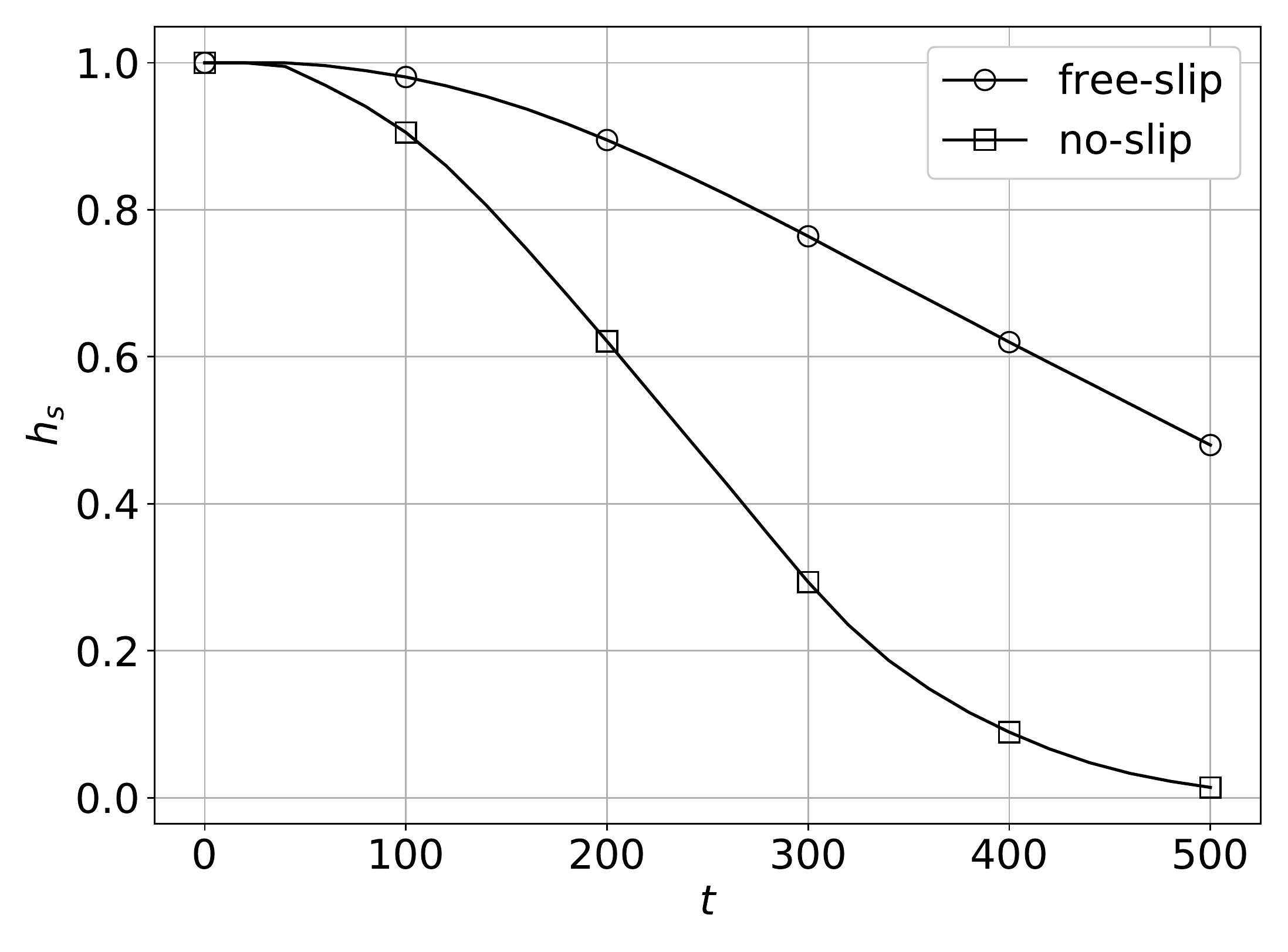}
\par\end{centering}
\caption{\label{fig:hs_noslip_BC} The melting histories with either a no-slip or a free-slip lower boundary.
The other parameters are identical, with $E=8\times10^{-5}$, $Ra=7.8\times10^6$, $Pr=5$, $St=1$, $f=0$.
Due to the enhanced heat transport, the rate of melting is higher with a no-slip lower boundary. }
\end{figure}

In the simulations presented thus far, the fluid layer is bounded
laterally and above by no-slip boundaries. Only the heated lower boundary
is one of free-slip. In rotating Rayleigh-B�nard convection, the role
of the velocity boundary layers is as essential as in the non-rotating
case \citep[e.g.,][]{Rossby1969, Liu:2009, Schmitz:2010,
Julien2012, King2012}. Morever, the critical Rayleigh number in Eq.
\ref{eq:Ra_cr_bulk} is largest for free-slip top- and bottom-boundaries,
and smallest for one free-slip and one no-slip boundary; the case of
two no-slip boundaries is intermediate between these cases \citep{Chandrasekhar1953,boubnov1990}.
Despite this, for the parameter ranges considered here, the Nusselt
number is larger for the case with no-slip upper and lower boundaries,
owing to the interaction of the thermal and velocity boundary layers
at the lower boundary \citep[e.g.,][]{Rossby1969}. Thus, the melting
rates are higher when the lower boundary is one of no-slip as compared to one of free-slip, as seen in Fig.
\ref{fig:hs_noslip_BC}.

\begin{figure}
\noindent \begin{centering}
\includegraphics[width=0.9\columnwidth]{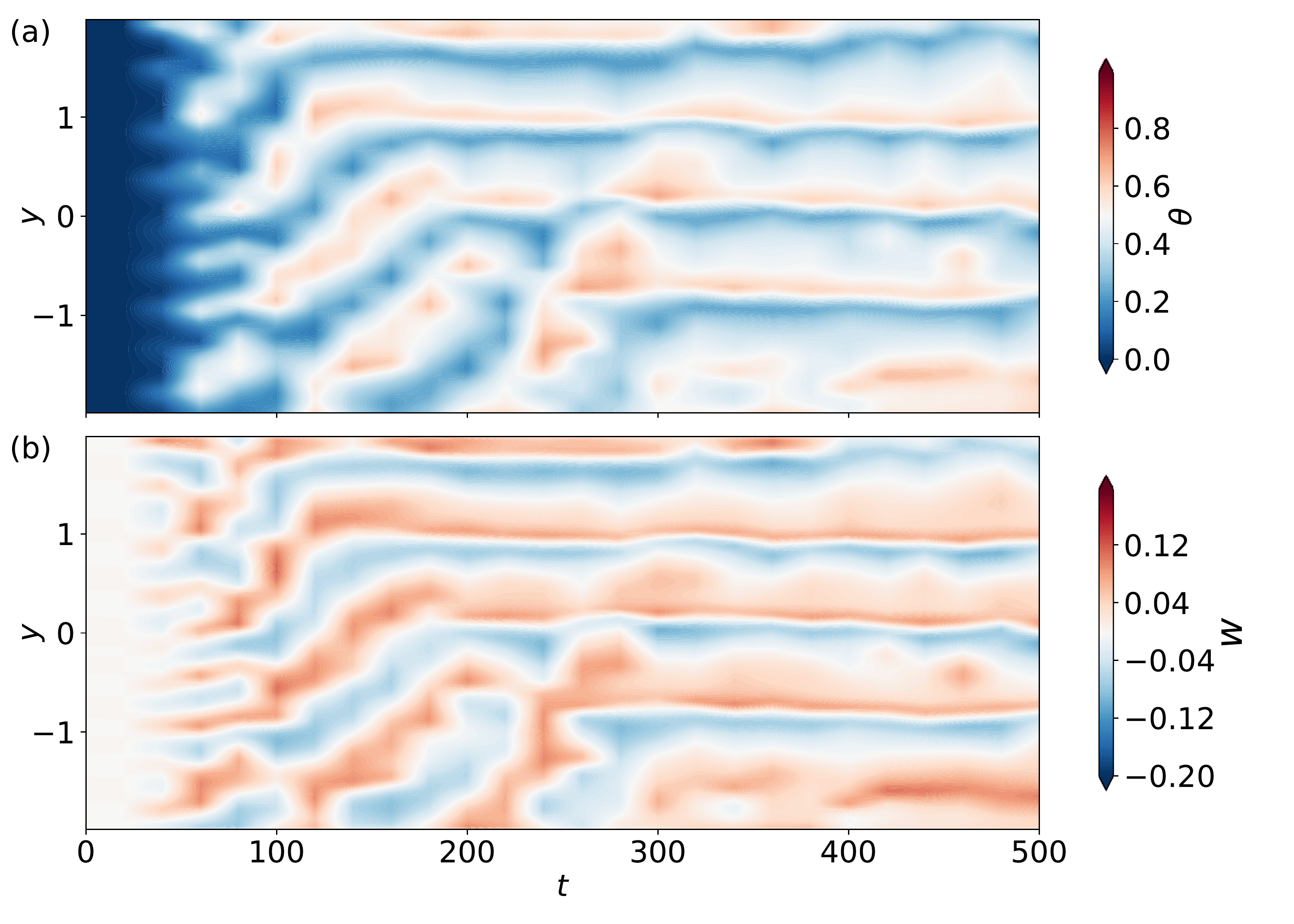}
\par\end{centering}
\caption{{\label{fig:hovmoller_noslip} A H\"ovm\"oller plot of the temperature, $\theta$, and the vertical velocity, w, at one of the vertical
walls. The wall-modes, which usually propagate clockwise, are locked in place once melting begins. The parameters are  
$E=8\times10^{-5}$, $Ra=1.56\times10^6$, $Pr=1$, $St=1$, $f=0$.}}
\end{figure}

{Experiments show that columnar vortices in rotating convection show horizontally diffusive motion \citep[see e.g.,][]{Noto2019}.
Because the phase boundary voids created by the heat transported through the columnar vortices are colocated, the latter can be 
arrested (and perhaps pinned) by the former.  In our simulations, this effect is influenced by velocity boundary conditions, with horizontal motion 
suppressed in the case of no-slip boundaries.  In Fig. \ref{fig:hovmoller_noslip}, we show that the wall-modes that usually precess in a retrograde (i.e. clockwise as seen from above) direction are locked in place as the solid melts, an effect that is more prominent with a no-slip lower
boundary than with a free-slip lower boundary.}

\subsubsection{Horizontal periodicity \label{subsec:Periodic}}

\begin{figure}
\includegraphics[width=0.48\columnwidth]{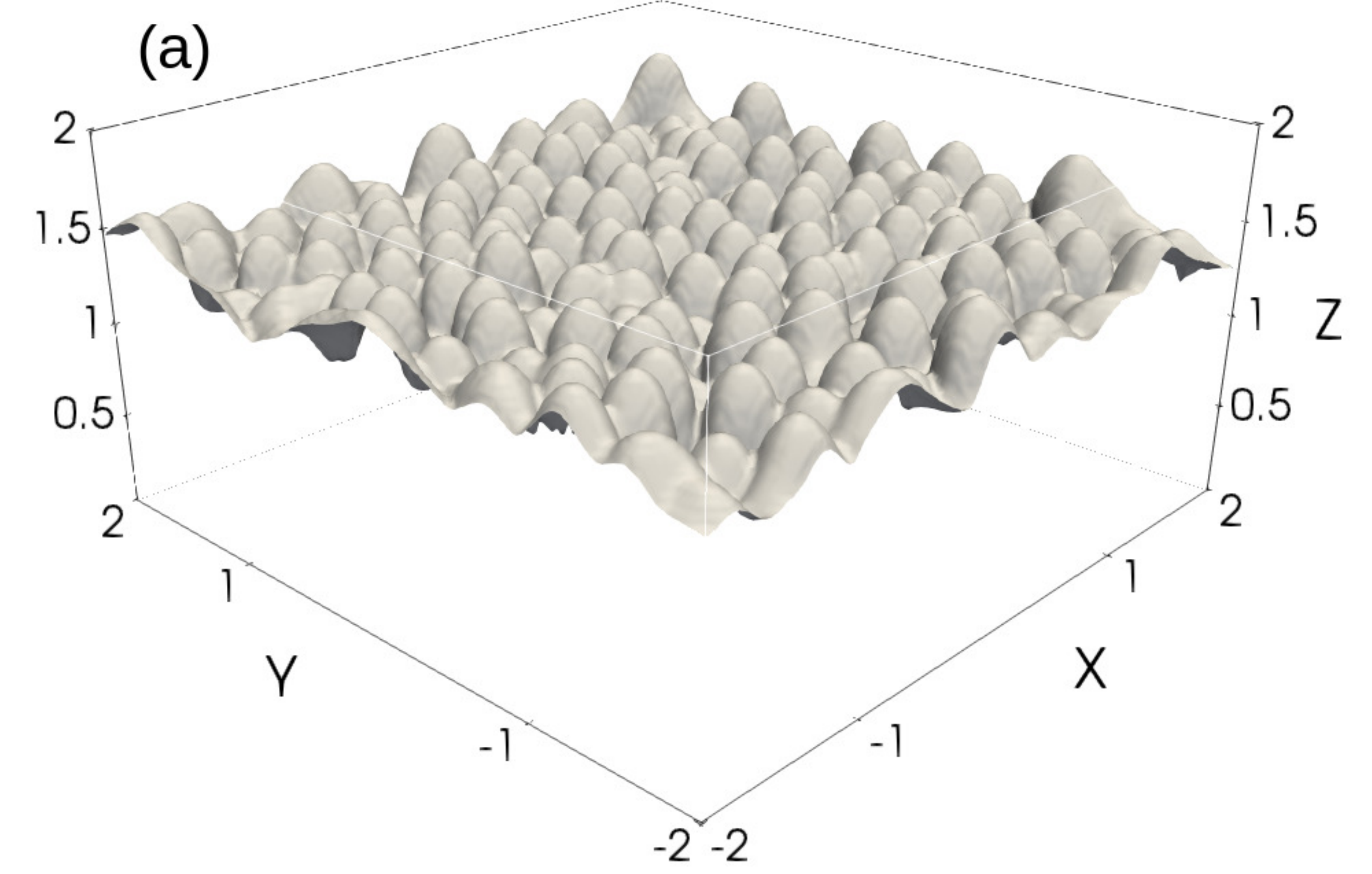}
\includegraphics[width=0.48\columnwidth]{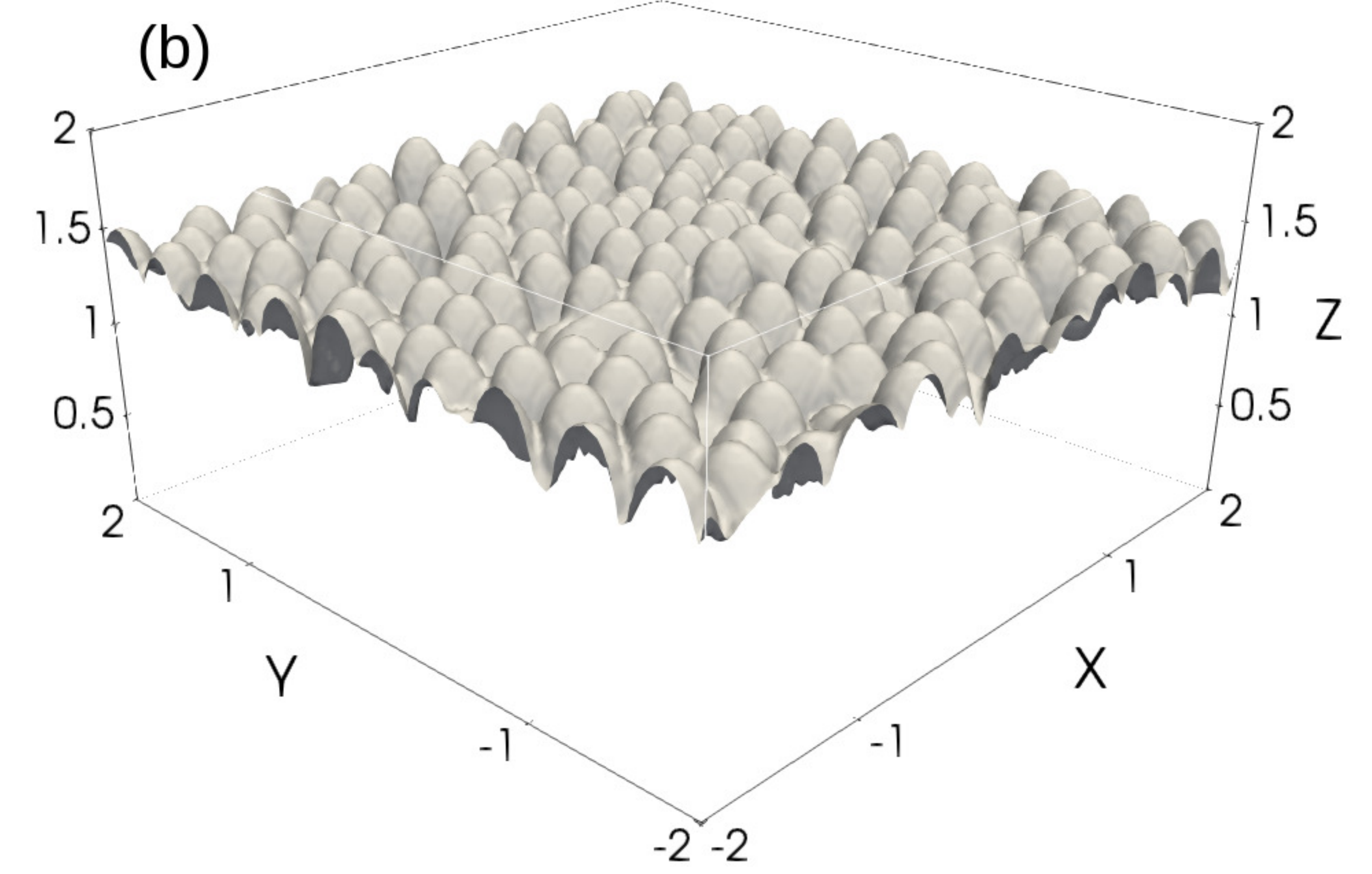}
\caption{\label{fig:interface_periodic_vs_walled}The solid-liquid interface
(viewed from the solid side) at {$t=400$} for {$E=8\times10^{-5}$,
$Ra=7.8\times10^{6}$}, $Pr=5$, $St=1$, {$f=0$}, and (a) no-slip walls (b)
periodic in the horizontal. The effects of the peripheral streaming
flow seen in (a) as increased melting near the walls, is absent in
(b), although the voids and the overall rate of melting are very similar
in the two cases.}
\end{figure}

As we have seen, the presence of walls confining the flow in the horizontal directions leads to the generation of a peripheral current that can
affect the melting of the solid. This peripheral flow is absent in a horizontally periodic system, as seen in Fig. \ref{fig:interface_periodic_vs_walled}.  However, the 
columnar-vortical flow at $Pr=5$ and the resultant melt pattern reflecting the presence of these vortices, as well as the overall melt rate, both remain
unchanged.

\subsubsection{Thermal diffusivity in the solid \label{subsec:vary_kappas}}

\begin{figure}
\includegraphics[width=0.43\columnwidth]{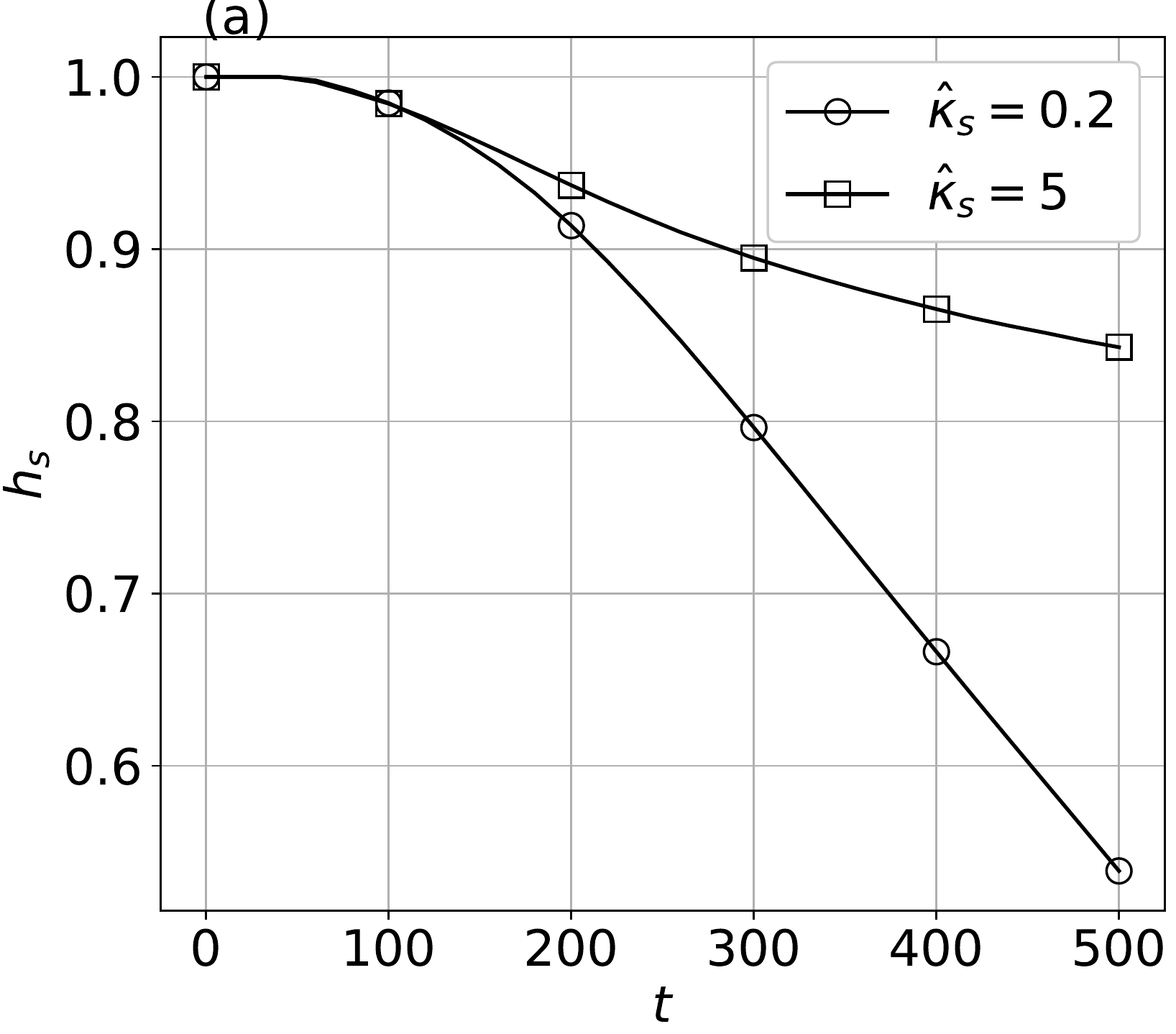}
\includegraphics[width=0.55\columnwidth]{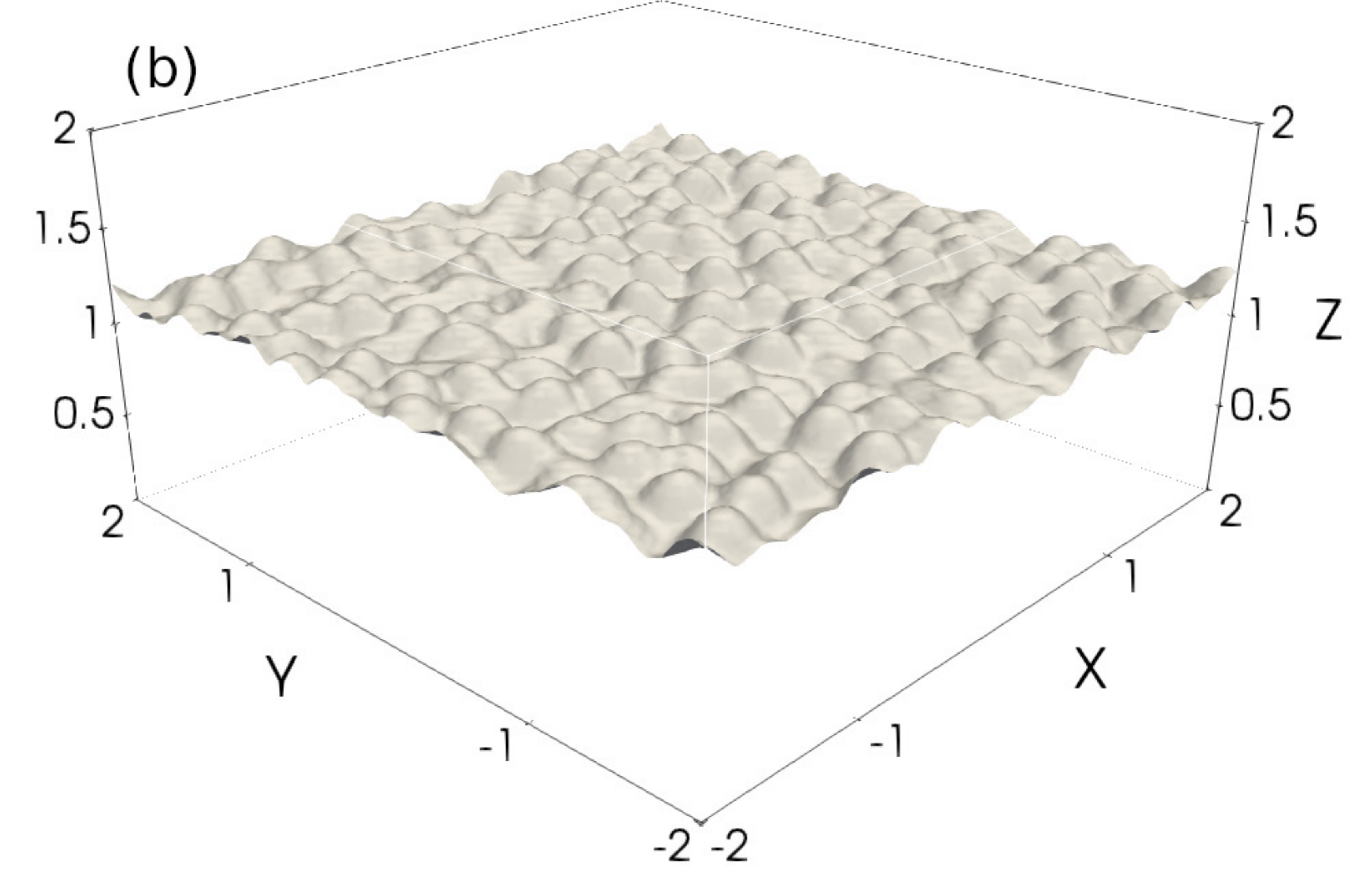}
\caption{\label{fig:vary_kappas} (a) The melting histories for $E=8\times10^{-5},$
$Ra=7.8\times10^{6}$, $St=1$, $f=1$, $Pr=5$ with periodic boundary
conditions in the horizontal for (i) $\hat{\kappa}_{s}=0.2$
and (ii) $\hat{\kappa}_{s}=5$. As $\hat{\kappa}_{s}$ increases the melt rate decreases.
(b) The solid-liquid interface (viewed from the solid side) for $\hat{\kappa}_{s}=5$ at {$t=500$}. The system is periodic in the horizontal, and the other parameters are as in (a), but the voids are not as prominent. }
\end{figure}

The thermal diffusivity of the solid governs the amount of heat transported
away from the solid-liquid interface and thus the melt rate (see Eq. \ref{eq:nondim_meltvel}),
with a larger diffusivity in the solid $\hat{\kappa}_{s}$ leading to smaller $u_m$.
Figure \ref{fig:vary_kappas}(a) shows this effect for two values of $\hat{\kappa}_{s}=0.2$ and $\hat{\kappa}_{s}=5$,
with $f=1$ (so that the upper boundary is at $\theta=-1$). For the largest value of the diffusivity, $\hat{\kappa}_{s}=5$, and the {smaller melting rate (see, Eq. \ref{eq:meltvel}),} the horizontal drift of the columnar vortices is faster than the melt rate and hence we infer the vortices are not pinned in the voids. 
{As a result, we see in Fig. \ref{fig:vary_kappas}(b) that the voids have smaller amplitudes.}

\begin{figure}
\includegraphics[width=1\columnwidth]{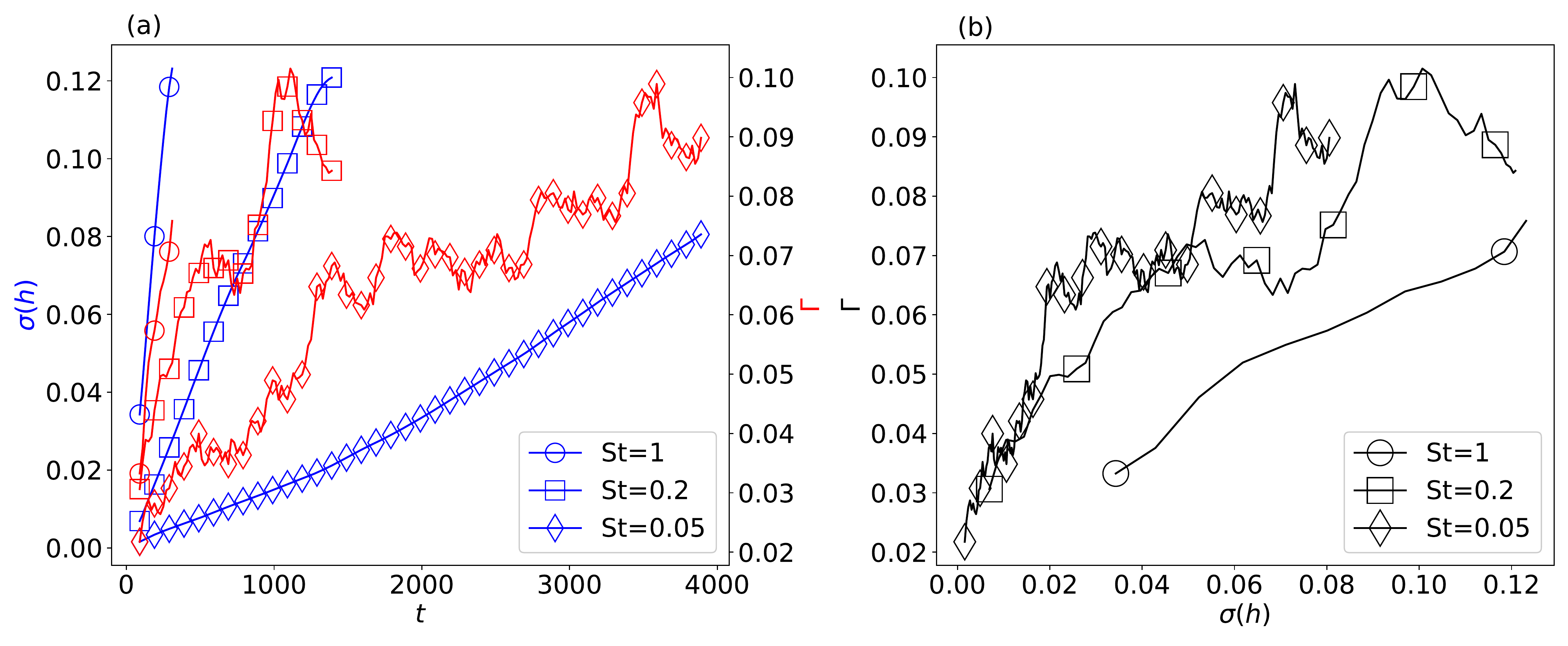}
\caption{\label{fig:vorticity_roughness} {(a) The net circulation ($\Gamma = \int \int \omega_z dx dy$) at $z=H/4$ and the roughness as a function of time; and (b) the net circulation $\Gamma$ as a function of the roughness, characterized by the standard deviation of the liquid height, $\sigma (h)$,  for simulations with $E=3.2\times10^{-4}$, $Ra=2\times10^{6}$, $f=0$, $Pr=5$ and $St=0.05, 0.2, 1$ showing that the total vorticity and the roughness co-vary. The curves are computed using a running average over $10$ points, each spaced $20$ flow units apart.}}
\end{figure}

\subsection{{Coupling of interfacial geometry and flow structure} \label{subsec:pinning}}
We argued in \S \ref{subsec:vary_kappas} that the phase boundary and the flow structures co-evolve, which is particularly well reflected in Fig. \ref{fig:nvor_vs_vorarea} 
showing the proportionality between the number and area of the vortices for $St=1$.  
Whilst we are unable to track individual vortices in our simulations, in Fig. \ref{fig:vorticity_roughness} we assess their interaction with the voids by plotting the time evolution of the 
net vertical circulation, or vorticity, and the roughness, as characterized by the standard deviation of the liquid height $\sigma (h)$.   We see that the rates at which both roughness and vorticity increase, decrease as the latent heat increases and that the roughness and the vorticity increase collinearly, which is a natural consequence of the conservation of potential vorticity.  Indeed, we speculate that the increase in vorticity with latent heat shown in Fig. \ref{fig:vorticity_roughness}(b) is associated with the horizontal drift of the columnar vortices being faster than the evolution of the phase boundary.  However, in order to assess such a scenario one must track individual vortices.

\subsection{Heat transport and the melting rate \label{subsec:melting_vs_Nusselt}}

\begin{figure}
\includegraphics[width=1\columnwidth]{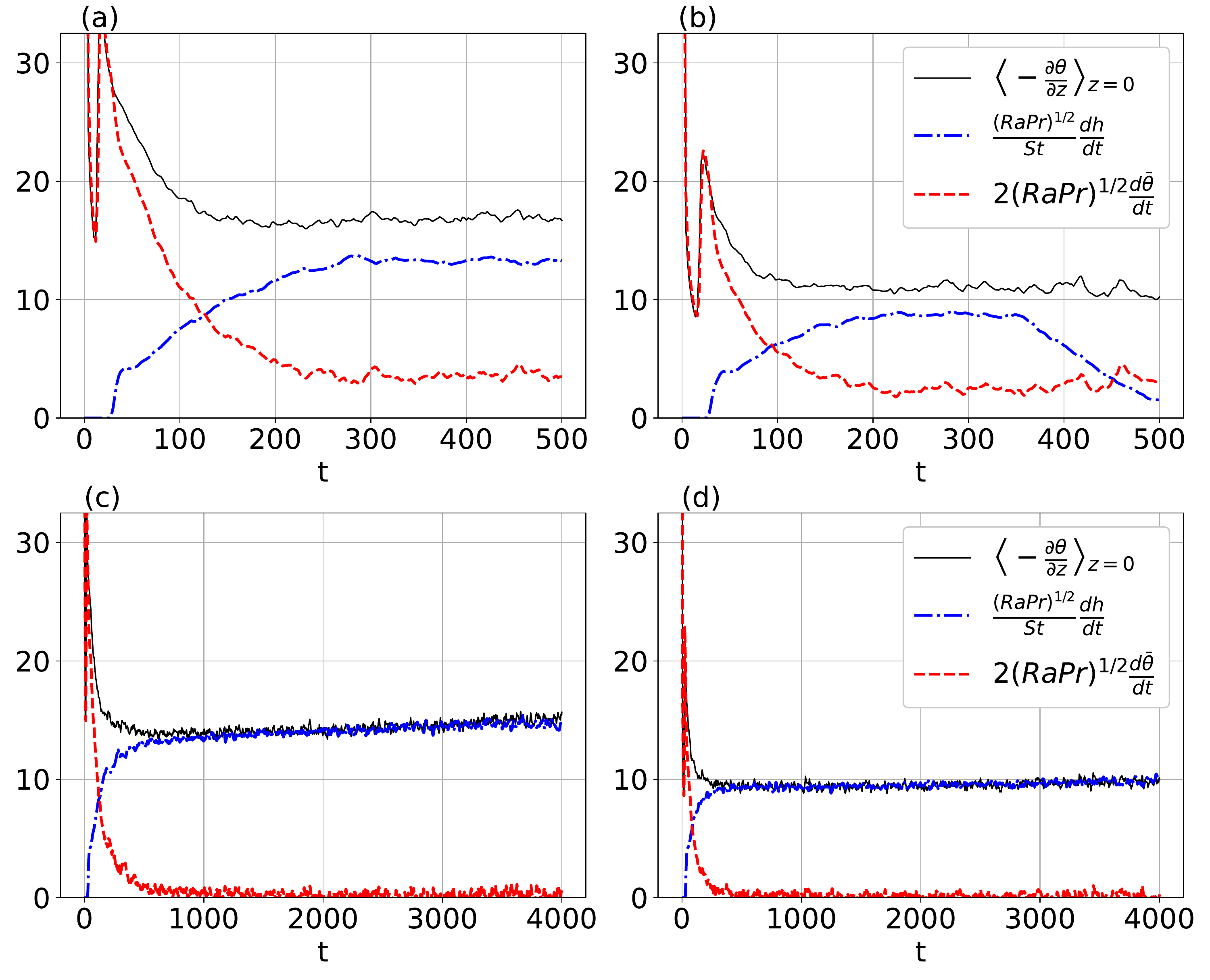}
\caption{\label{fig:heat_balance} The terms in Eq. \ref{eq:heat_balance},
with $Pr=5$ and {$f=0$} for (a,c) {$E=10^{-4}$ $Ra=10^{7}$}; (b,d) {$E=3.2\times10^{-4}$,
$Ra=2\times10^{6}$}. The Stefan numbers are (a,b) $St=1$; (c) $St=0.1$;
(d) $St=0.05$. Note that the quasi-steady state of convection in
the fluid described by Eq. \ref{eq:heat_balance} breaks down when
the voids in the solid reach the upper boundary and fluid comes into
direct contact with the container surface at {$t=340$} in (b). }
\end{figure}

In \S \ref{subsec:ICs_BCs} we noted that the initial and boundary
conditions in most of the simulations reported here, except those
in \S\ref{subsec:vary_kappas}, are that the solid is at the melting
temperature throughout, viz., $\theta(t=0)=0$, and the upper boundary
is held at $\theta=0$. Therefore, the heat available for melting
is transported by the fluid from the lower heated boundary to the
solid and described by the integral form of energy conservation, Eq.
\ref{eq:enthalpy}, as 
\begin{widetext}
\begin{align*}
\rho\lambda (H/2)^2 U_{b}\left[\frac{d}{dt}\iiint(1-\chi)dxdydz\right]
& = k_{l}\Delta T A^2 (H/2)\left[\left\langle -\frac{\partial\theta}{\partial z}\right\rangle _{z=0}\right] - \rho C_{p}\Delta T (H/2)^{2}U_{b}\left[\frac{d}{dt}\iiint\theta dxdydz\right],
\end{align*}
\end{widetext}
where the terms in square brackets are nondimensional. Dividing 
by $k_{l}\Delta T A^2 H/2$ gives 
\begin{equation}
{\frac{\left(Ra Pr\right)^{1/2}}{St}\frac{dh}{dt}=\left\langle -\frac{\partial\theta}{\partial z}\right\rangle _{z=0}- 2\left(Ra Pr\right)^{1/2}\frac{d\bar{\theta}}{dt},}
\label{eq:heat_balance}
\end{equation}
where 
\begin{equation}
{\bar{\theta}=\frac{1}{2A^2}\iiint\theta dxdydz\label{eq:theta_bar}}
\end{equation}
is the average nondimensional temperature over the simulation volume
and 
\begin{equation}
{h=\frac{1}{A^2}\iiint(1-\chi)dxdydz\label{eq:h_defn}}
\end{equation}
is the volume-averaged dimensionless height of the fluid. The relative
contributions of the sensible heating of the fluid and the melting
of the solid to the heat balance are shown in Fig. \ref{fig:heat_balance}.
Initially, all the energy supplied to the system from the boundary
heats up the liquid. For {smaller $E$ and $Ra$,} vertical motions are
suppressed and hence so too is the delivery of the specific heat to
the phase boundary, where melting may proceed (beginning here at about
{$t=50$}). Once melting begins the latent heat draws down the sensible
heat stored in the fluid and eventually a near steady balance between
the energy delivered and that available for melting may be maintained.
Hence, whilst the vigor of convection depends {on $E$ and $Ra$}, such
a balance {between the heat input at the lower boundary and the
latent heat of fusion} requires quasi-steady rotating convection.

We see in Fig. \ref{fig:heat_balance}(b) that the quasi-steady state
of convection in the fluid described by Eq. \ref{eq:heat_balance}
breaks down at {$t=340$} when fluid comes into contact with the upper
solid boundary through the voids in the solid. Note that the slight
mismatch between $\left<-\frac{\partial\theta}{\partial z}\right>_{z=0}$
and the sum {$\frac{\left(Ra Pr\right)^{1/2}}{St}\frac{dh}{dt}+2\left(Ra Pr\right)^{1/2}\frac{d\bar{\theta}}{dt}$}
in Figs. \ref{fig:heat_balance} is a consequence of the coarse time-discretization
used in calculating the time-derivatives in the plots.

\begin{figure}
\begin{centering}
\includegraphics[width=0.5\columnwidth]{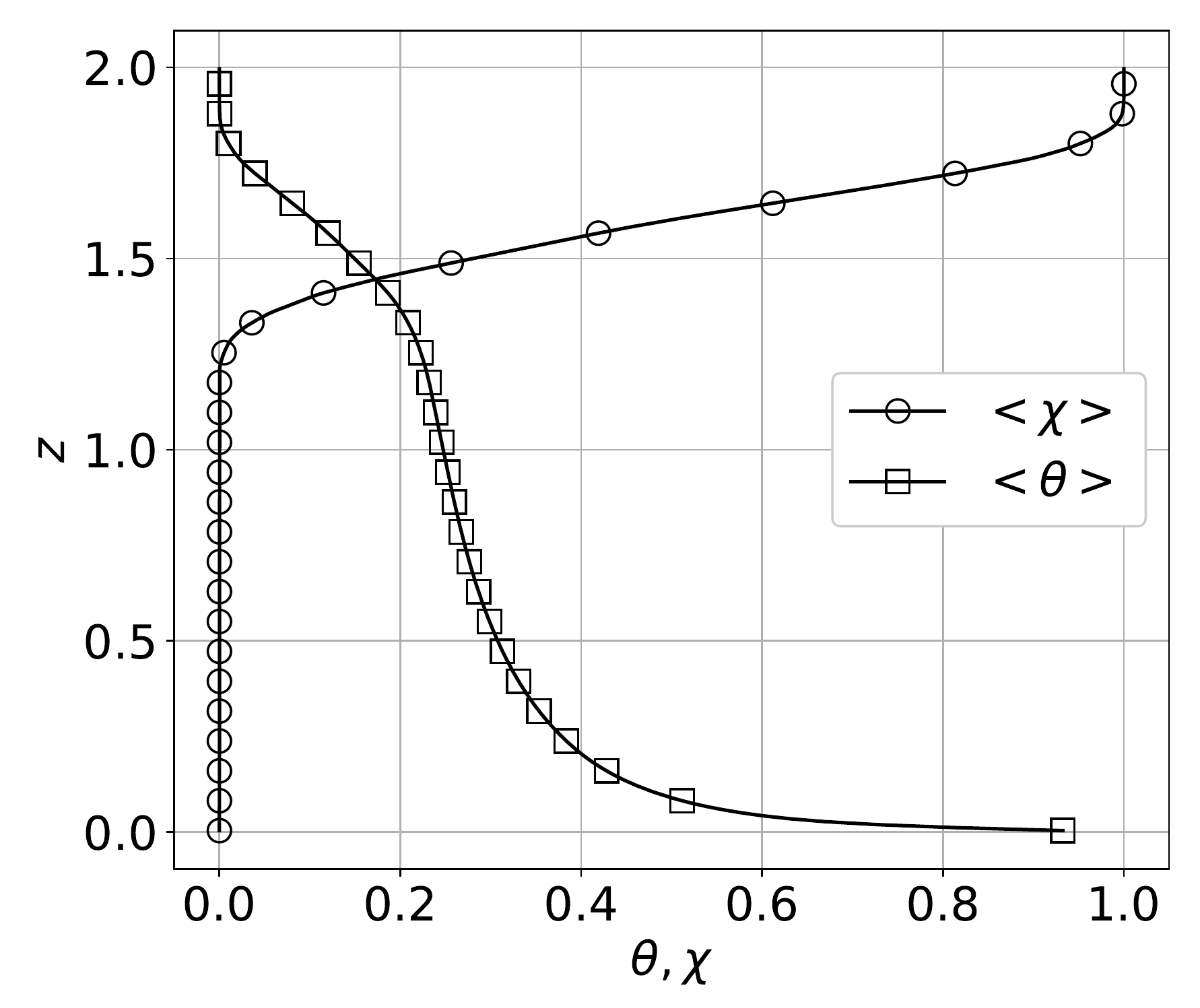}
\par\end{centering}
\caption{\label{fig:theta_vs_z} The area-averaged solid fraction $\langle\chi\rangle$
and temperature $\langle\theta\rangle$ as a function of the vertical
coordinate $z$ at {$t=400$}, for the case {$E=10^{-4}$, $Ra=10^{7}$,}
$Pr=5$, $St=1$ and $f=0$.}
\end{figure}

Additionally Fig. \ref{fig:heat_balance} shows that when the specific
heat stored in the convecting fluid is small, i.e. when the Stefan
number is small, there is a nearly steady balance between the heat
supplied at the base of the cell and the melt rate.
As the fluid interior cools slightly in time this is balanced by a slight 
increase of the melt rate and the heat input from the lower boundary, as
seen in Figs. \ref{fig:heat_balance}(c) and (d).
The temperature in the liquid is, of course,
not uniform in space. Indeed, as shown in Fig. \ref{fig:theta_vs_z},
the structure of the mean temperature gradient in the fluid is reminiscent
of non-rotating high $Ra$ convection, with a thermal boundary layer
at the base and a nearly isothermal interior. However, the phase change
at the ramified upper boundary maintains the average temperature near
the melting point. This situation can be treated by approximating
Eq. \ref{eq:heat_balance} using only the first two terms, viz., 
\begin{align}
{\frac{\left(Ra Pr\right)^{1/2}}{St}\frac{dh}{dt}=\frac{\text{Nu}}{h},\label{eq:melting_rate}}
\end{align}
where Nu is the Nusselt number--the total heat flux scaled
by the conductive heat flux--across the fluid region. 

In \S \ref{subsec:Flow-structures} we showed that for most combinations
of parameters examined here, the phase boundary is ramified, so that
the solid depth varies substantially in the horizontal. In
consequence, we see from Fig. \ref{fig:theta_vs_z} that within the broad
average transition region from fluid to solid the average temperature
relaxes to the bulk melting temperature. Therefore, we take the domain
averaged $h$ \citep[see also Section III of][]{RabbanipourEsfahani2018}
when considering the quasi-steady balance in Eq. \ref{eq:melting_rate}.
We note, however, that we understand that there are three-dimensional
heat fluxes in the interfacial region, which are simpler to treat when
the phase boundary has small amplitude variations, such as in the
non-rotating case \citep[e.g.,][]{Favier2019, Toppaladoddi2019}. Another
perspective is that for a vortex-induced highly ramified interface,
the interfacial region might be considered as a ``mushy layer'', as
observed in binary systems \citep{worster2000solidification}, wherein
there is two-phase, two-component coexistence and the condition of
marginal equilibrium holds. Clearly here there are no impurities,
but we can see in Fig. \ref{fig:theta_vs_z} the relaxation towards
equilibrium of the average temperature and enthalpy through the mixed
phase region.

For geostrophic convection, the average Nu can be expressed in terms
of the Rayleigh number and the critical Rayleigh number, using Eqs.
(\ref{eq:Ra_defn}), (\ref{eq:Ekman_defn}) and (\ref{eq:Ra_cr_bulk}),
as 
\begin{equation}
\text{Nu}=C\left(\frac{Ra}{Ra_{c}^{\text{bulk}}}\right)_{\text{eff}}^{\beta},\label{eq:Nu_vs_Ra_by_Rac}
\end{equation}
where $\beta$ is in general a function of $\left(Ra/Ra_{c}^{\text{bulk}}\right)_{\text{eff}}$
and C is a numerical prefactor that may depend on $Pr$. For large
values of $Ra/Ra_{c}^{\text{bulk}}$, two values have been suggested
in the literature; $\beta=3$ \citep{boubnov1990,King2012} and $\beta=3/2$
\citep{Julien2012}, the latter finding C=$(1/25)Pr^{-1/2}$. For
more modest values of $Ra/Ra_{c}^{\text{bulk}}$, \cite{Ravichandran2020_RotCon}
found $\beta=3/4$ and \cite{Liu:2009} found $\beta=2/7$. In the
limit of large $Ro$, that is in the classical non-rotating Rayleigh-B�nard
convection regime, one finds, with a different prefactor than in Eq.
\ref{eq:Nu_vs_Ra_by_Rac}, $\beta=1/3$ up to $Ra=10^{15}$ \citep{DTW:2019, Doering:2020a, Doering:2020b, Iyer:2020}.

\begin{figure}
\begin{centering}
\includegraphics[width=0.8\columnwidth]{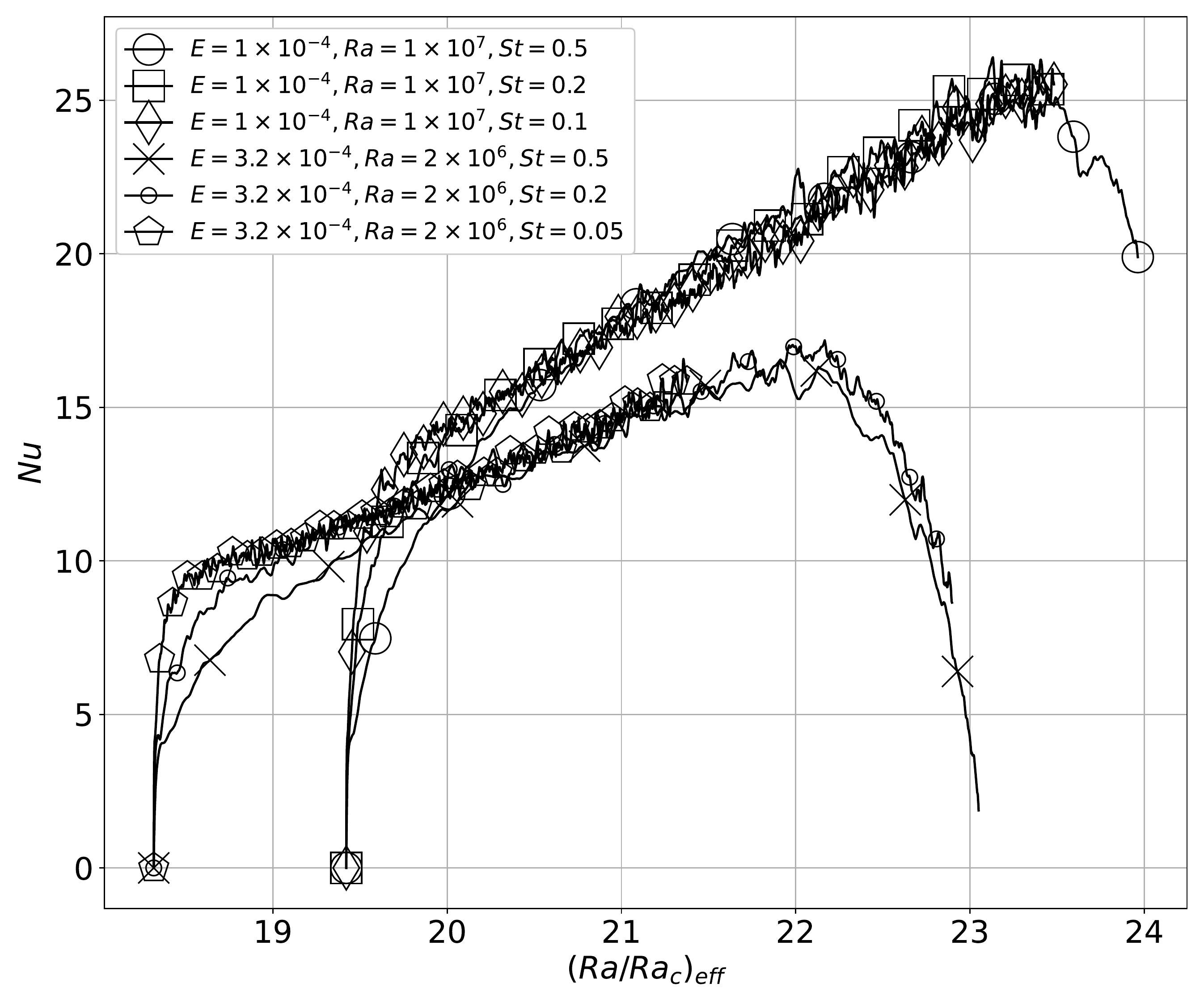}
\par\end{centering}
\caption{\label{fig:Nu_vs_Raeff} The Nusselt number plotted as a function
of $\left(Ra/Ra_{c}\right)_{\text{eff}}$, calculated using the mean
fluid height $h(t)$ and Eqs. (\ref{eq:Ra_defn}) and (\ref{eq:Ekman_defn}).}
\end{figure}

In Fig. \ref{fig:Nu_vs_Raeff} we plot the Nusselt number, calculated using Eq. (\ref{eq:melting_rate}),
versus the effective Rayleigh number as melting proceeds. In the quasi-steady state the curves for different $St$ collapse with {$E$ and $Ra$ dependent slopes}, suggesting that although Eq. (\ref{eq:Nu_vs_Ra_by_Rac}) provides an ideal organizing principle for our simulations, we are unable to determine the associated exponent given our parameter range \cite[see e.g.,][]{Stumpf2012}.

\subsection{Maximal Phase Boundary Roughness \& Maximal Heat Flux \label{sec:optimal}}

\begin{figure}
\begin{centering}
\includegraphics[width=1\columnwidth]{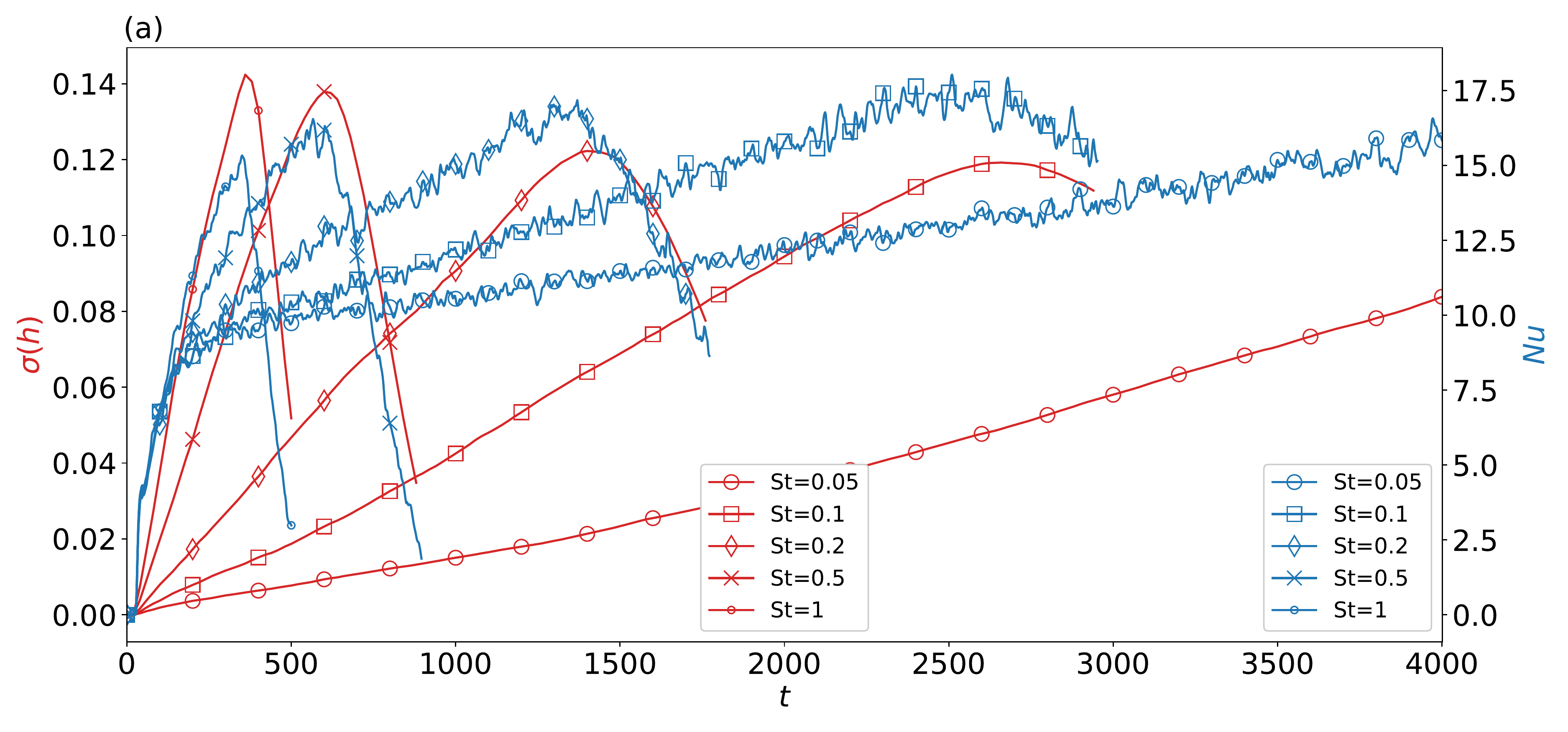}
\par\end{centering}
\begin{centering}
\includegraphics[width=1\columnwidth]{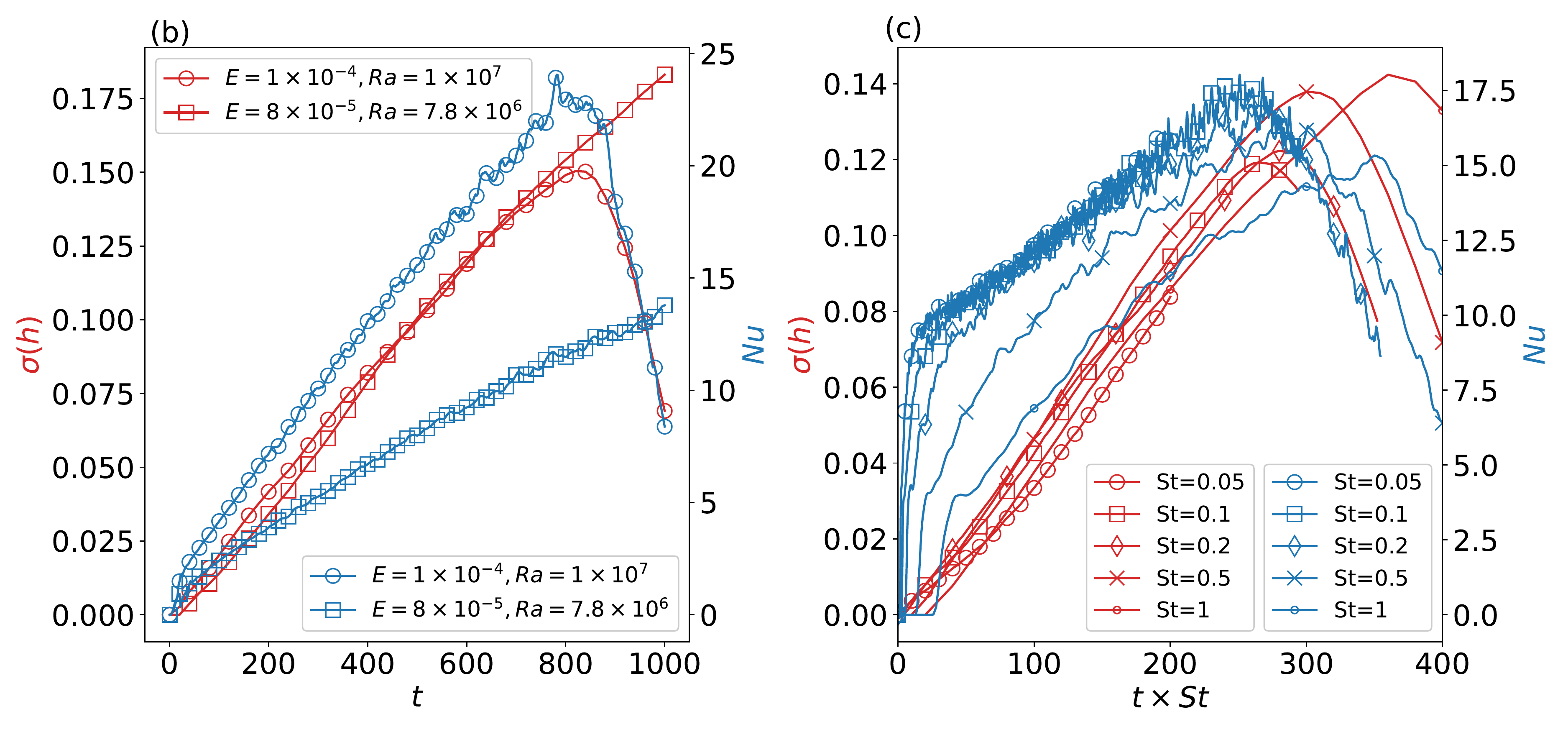}
\par\end{centering}
\caption{\label{fig:Nu_sigma_h} The Nusselt number Nu and standard deviation of the phase boundary height, $\sigma(h)$, plotted as a function of time for {$f=0$} and (a) $E=3.2\times10^{-4}$, $Ra=2\times10^6$, $Pr=5$, {$h_{0}=1.0$,}
for a range of $St$; and for (b) $St=1$ with two combinations of
$E,Ra$ and {$h_{0}=0.1$}. The correlation between Nu and $\sigma\left(h\right)$
is evident in all of these cases. In (c), we show  the roughness data in (a) with the time coordinate rescaled by the Stefan number.  Thus, we see the Nusselt number maxima occurring at smaller $t \times St$ for smaller $St$, from which we expect that the data for $St=0.05$ will follow this trend and reach a maximum, were we able to run longer simulations in that case. }
\end{figure}

We conclude \S \ref{sec:Results} with the observation that the roughness of the phase boundary continuously increases and reaches a maximum approximately simultaneously with the Nusselt number.  As seen in Fig. \ref{fig:heat_balance}, the heat supplied at the bottom boundary, and the melt rate of the solid, are approximately independent of time and hence the left hand side of Eq. \ref{eq:melting_rate} is approximately constant.  Therefore, the Nusselt number increases linearly with the liquid height $h$ and reaches a maximum when the voids in the solid reach the upper boundary of the cell. 
We again characterize the roughness using the standard deviation of the liquid height, $\sigma (h)$, which we observe reaches a maximal value  when the voids reach the upper boundary, namely when there is fluid in contact with the upper boundary. Further melting reduces the roughness. The correlation between Nu and $\sigma (h)$ is shown in Fig. \ref{fig:Nu_sigma_h}(a), where we see that the maximal Nusselt numbers are reached before the roughness of the solid-liquid interface becomes maximal, with the interval between the maxima increasing as the Stefan number decreases (and the melt rate decreases). Smaller Stefan numbers lead to voids of unequal depths, with some voids reaching the upper boundary before others. The decrease of the interface roughness associated with the former is compensated, for a limited period, by the continued deepening of the latter.

Since the areas and number of voids depend on the
flow parameters (Fig. \ref{fig:nvors_vs_nholes}), the maximum value
of $\sigma\left(h\right)$ depends on these parameters as well, with
the thinner vortices in flows with smaller $Ro_{c}$ (Eq. \ref{eq:Ro_convective})
leading to narrower voids and thus a rougher interface (see e.g., 
Fig.\ref{fig:Nu_sigma_h}(b)). In particular, the continued increase
of $\sigma\left(h\right)$ in Fig. \ref{fig:Nu_sigma_h}(b), where
the initial liquid height {$h_{0}=0.1$}, shows that the voids formed
by the columnar vortices will continue to penetrate deeper into the
solid with time, only being limited by the depth of the solid itself.
The curves for $St=0.05$ in Fig. \ref{fig:Nu_sigma_h}(a) have not reached their
maxima. However, the rescaling of the data in Fig. \ref{fig:Nu_sigma_h}(c) suggests that
the time interval between the maximal Nu and the maximal $\sigma (h)$ will further increase 
for $St=0.05$, and we expect that a maximum will be reached were we able to run longer simulations in that case.

In non-rotating turbulent Rayleigh-B�nard convection, with Dirichlet
boundary conditions and periodically rough boundaries, \cite{Toppaladoddi:2015,Toppaladoddi2017}
showed that, for a given roughness wavelength, there is a ratio of
the thermal boundary layer thickness to the roughness amplitude that
optimizes Nu.  Moreover, this enhancement of heat transport  is a general consequence of roughness,  
observed for a wide range of geometries from rough on all surfaces to fractal boundaries \cite[]{Roche, GoluskinDoering, TWDW2020}.
In these situations, the systems are in a statistical steady
state. Here, with the geometry free to evolve subject only to 
the underlying conservation laws, both the roughness of the phase boundary and the Nusselt number increase with time
as the solid phase melts according to Eq. \ref{eq:melting_rate}.  {The Stefan number dependence of the observed correlation between the vorticity and the interfacial roughness 
shown in Fig. \ref{fig:vorticity_roughness} underlies this process. }

\section{Conclusion \label{sec:Conclusions}}

We have studied the melting of a pure solid by the convection of its liquid phase
when the former overlies the latter and the entire system rotates about an axis parallel to gravity. The
width of the system is twice its depth and we have examined ranges
of the Ekman, Rayleigh and Prandtl numbers predominantly corresponding to moderately
rotating Rayleigh-B\'{e}nard convection.

There are three regimes of flow that influence the morphology of the
phase boundary. First, when the Rayleigh number is greater than the
bulk critical value, $Ra>Ra_{c}^{\text{bulk}}$ (Eq. \ref{eq:Ra_cr_bulk}),
the flow takes the form of columnar vortices. Second, in confined
geometries there is a streaming flow close to the lateral walls of
the container. This occurs when $Ra_{c}^{\text{wall}}<Ra<Ra_{c}^{\text{bulk}}$,
where $Ra_{c}^{\text{wall}}$ is given by Eq. \ref{eq:Ra_cr_wall}
\citep{Herrmann1993}. Third, in the periodic geometry, there is no
flow for $Ra<Ra_{c}^{\text{bulk}}$. We found that the number of melt
voids in the solid is proportional to the number of heat transporting
vortices present, which in turn increases as the convective Rossby number decreases and rotational effects become dominant.
We showed that the overall melting rate is a nontrivial function of the flow parameters; for the same $Ra/Ra_c$, 
melting rates are smaller for larger Prandtl numbers and smaller E.
Moreover, we found that the phase boundary morphology can be highly ramified 
or relatively smooth, reflecting the nature and number of rotationally 
controlled vortices transporting heat across the evolving fluid layer. 
{Lastly, we showed that the peripheral streaming current characteristic of rotating Rayleigh-B\'enard convection may become ``locked'' in place due to the coupling between the flow and the melting of the solid. }

For large values of the latent heat of fusion, characterized by the Stefan number, we found a quasi-steady
geostrophic convective state in which the net vertical heat flux is nearly constant over long
time intervals. This leads to a situation in which the constant heat supplied at the base balances the melt rate.
In the case of non-rotating binary systems, it is now well known that
the fluid mechanics of solidification lead to complex phase boundary
geometries and their associated transport phenomena \cite[e.g.,][]{HEH:1990,Sullivan:1996,worster2000solidification,
Philippi:2019}. 
Here in contrast, in a pure system, we find that convective and rotationally controlled vortices alone can create ramified phase
boundaries. While no obvious optimization of the Nusselt number is seen as a consequence
of the increasing boundary roughness, that  roughness evolves in time in a unique manner coupled to the rotationally influenced evolving buoyancy of the liquid phase.  
The associated void structure in the solid will affect the mechanical and thermal properties of materials formed in such circumstances.   
Thus, the inclusion of compositional effects with the rotational processes studied here will open a new set of questions regarding 
the structure of partially molten rotating systems.  Finally, we note that in astrophysical and geophysical problems wherein rotational
effects are important, assumptions of planarity of the phase boundary should
therefore be made with care.

\section*{Acknowledgements}

Computational resources from the Swedish National Infrastructure for
Computing (SNIC) under grants SNIC/2018-3-580, SNIC/2019-3-386 and
SNIC/2020-5-471 are gratefully acknowledged. Computations were performed
on Tetralith. The Swedish Research Council under grant no. 638-2013-9243,
is gratefully acknowledged for support.

\section*{Declaration of interests}

The authors report no conflict of interest.

\section*{Appendix A : Validation of the enthalpy method\label{sec:Appendix_Validation_enthalpy_method}}

We validate the enthalpy method used here by comparing the numerical solution to
the one-dimensional analytical solution for a purely conducting case
\cite[e.g.,][]{worster2000solidification}. We then study the convergence
of the method with grid resolution in a case with fluid convection.

\subsection*{Melting by conductive heat transfer} \label{subsec:validation_no_convection}
\begin{figure}
\includegraphics[width=1\columnwidth]{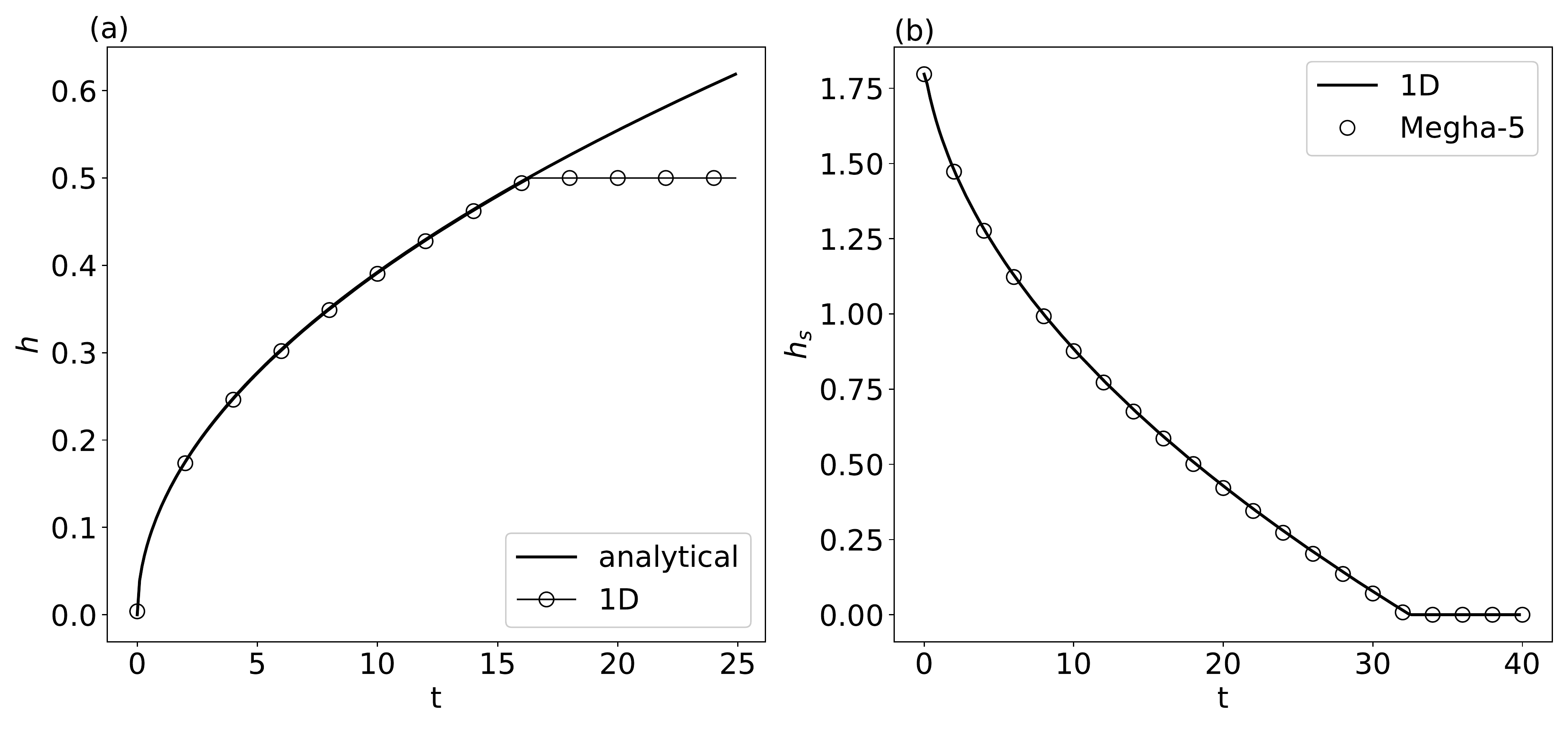}

\caption{\label{fig:validation_enthalpy_method}(a) The liquid height $h(t)$
from the one-dimensional numerical solution and the analytical solution
of the Stefan problem (Eq. \ref{eq:analytical_soln}). In the numerical
solution $h(t)$ is bounded by the height of the domain, $0.5$. (b)
For the alternate initial conditions (see text), the amount of unmelted
solid from the numerical solution from the finite-volume solver is
compared with the analytical solution in one dimension. The parameters
are $\kappa=0.01$, $St=1$.}
\end{figure}

Consider a semi-infinite solid layer in the region $z>0$ at the melting temperature.
The boundary at $z=0$ is held at $\theta=1$. The solid melts, forming
a liquid layer of height $h(t)$ given by 
\begin{equation}
h=2\xi\sqrt{\kappa t},\label{eq:analytical_soln}
\end{equation}
where $\xi$ is the solution of the transcendental equation deriving
from the Stefan condition, 
\[
\xi\text{exp}(\xi^{2})\text{erf}(\xi)=\frac{St}{\sqrt{\pi}}.
\]
In Fig. \ref{fig:validation_enthalpy_method} (a) the analytical solution
of the Stefan problem is compared with a numerical solution of Eq.
\ref{eq:enthalpy} in one dimension with the boundaries at $z=0$
and $z=H=0.5$. Next, we consider a case where there is already some
liquid (at $\theta=0$) present in the region $0<z<z_{0}=0.05$, with
solid at the melting temperature in the region $z_{0}<z<H$ at $\theta=0$.
The boundaries are held at $\theta(z=0)=1$ and $\theta(z=H)=0$.
The numerical solution in one dimension is compared with the solution
from the 3D solver, and the amount of unmelted solid plotted as a
function of time in Fig. \ref{fig:validation_enthalpy_method}(b).
In both these cases, the 1D solution is obtained using fourth-order
Runge-Kutta integration; the 3D solver uses a second-order Adams-Bashforth
scheme (as described in \S \ref{sec:Numerical-Simulations}).

\begin{figure}
\includegraphics[width=1\columnwidth]{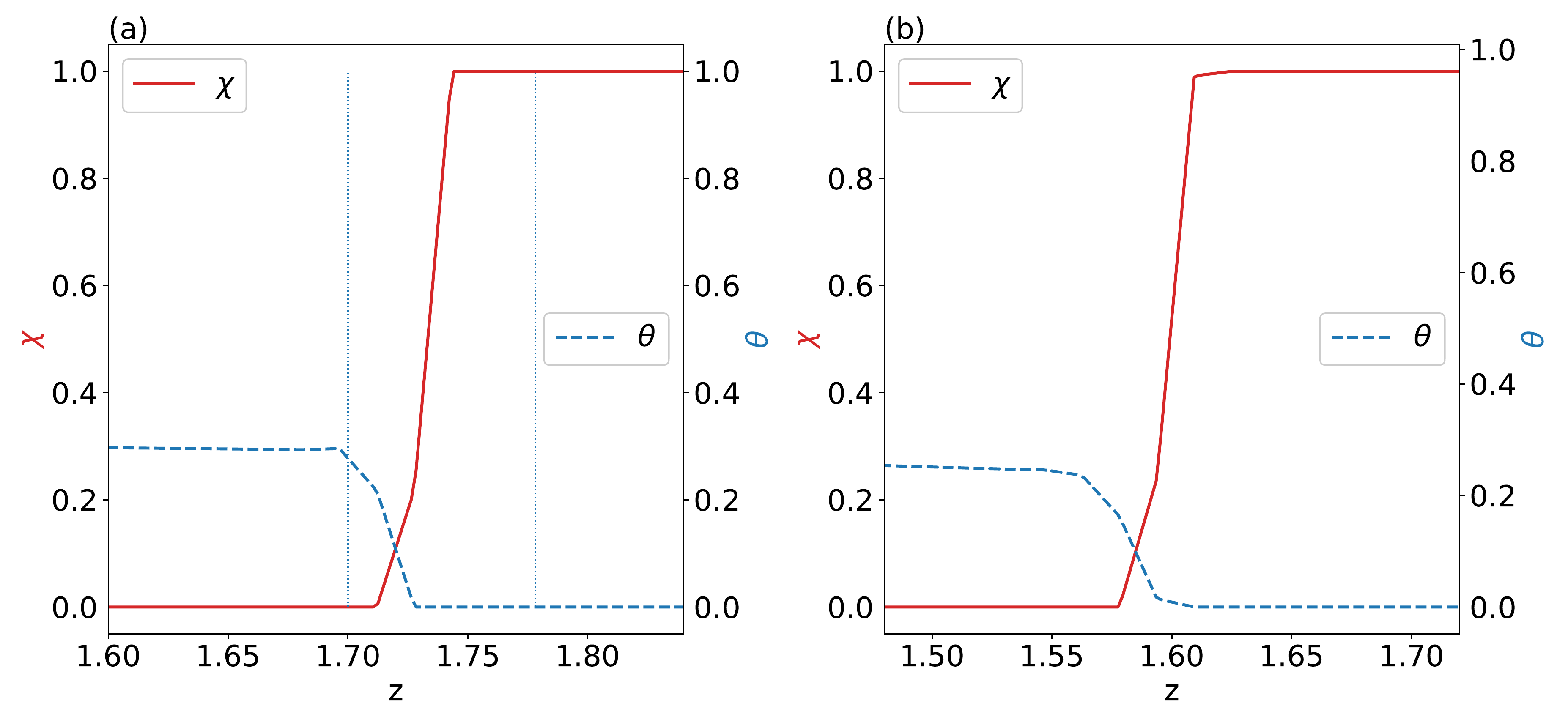}
\caption{\label{fig:chi_theta_vs_z}The variation of the solid mask and the
temperature through the solid-liquid interface. Parameters: $E=8\times10^{-5},$
$Ra=7.8\times10^{6}$, $Pr=5$, $St=1$, {$f=0$}, grid spacing $dz\approx0.004$
with (a) {$\eta=2\times10^{-3}$}, (b) {$\eta=10^{-3}$}. The dotted lines in (a) show
the average thickness of the thermal boundary layer at the heated lower boundary.}
\end{figure}

For the single-component, two-phase systems considered here, the solid-liquid
interface is sharp. In the numerical simulations, this interface is
defined as the region where $0<\chi<1$, and is distributed over a
finite number of gridpoints. This is shown in Fig. \ref{fig:chi_theta_vs_z}
where the the mask $\chi$ and the temperature $\theta$ are plotted
on a vertical line through the peak of the void in the solid region.
The mask function $\chi$ varies from $0$ to $1$ over a distance
of about $\delta z=0.008$, which is $2$ gridpoints in the $256^{2}\times128$
simulations. This is similar to results obtained by \cite{Couston2020}, and those prescribed (in their formulation) by \cite{Favier2019} who use a nominal interface thickness of half the grid-spacing.
We note that for the range of values of $\eta$ used here, the thinness of the interface is not affected
by changes in the grid-resolution or in the penalization parameter,
as seen from Fig. \ref{fig:chi_theta_vs_z}(b), with {$\eta=10^{-3}$}.

\begin{figure}
\includegraphics[width=1\columnwidth]{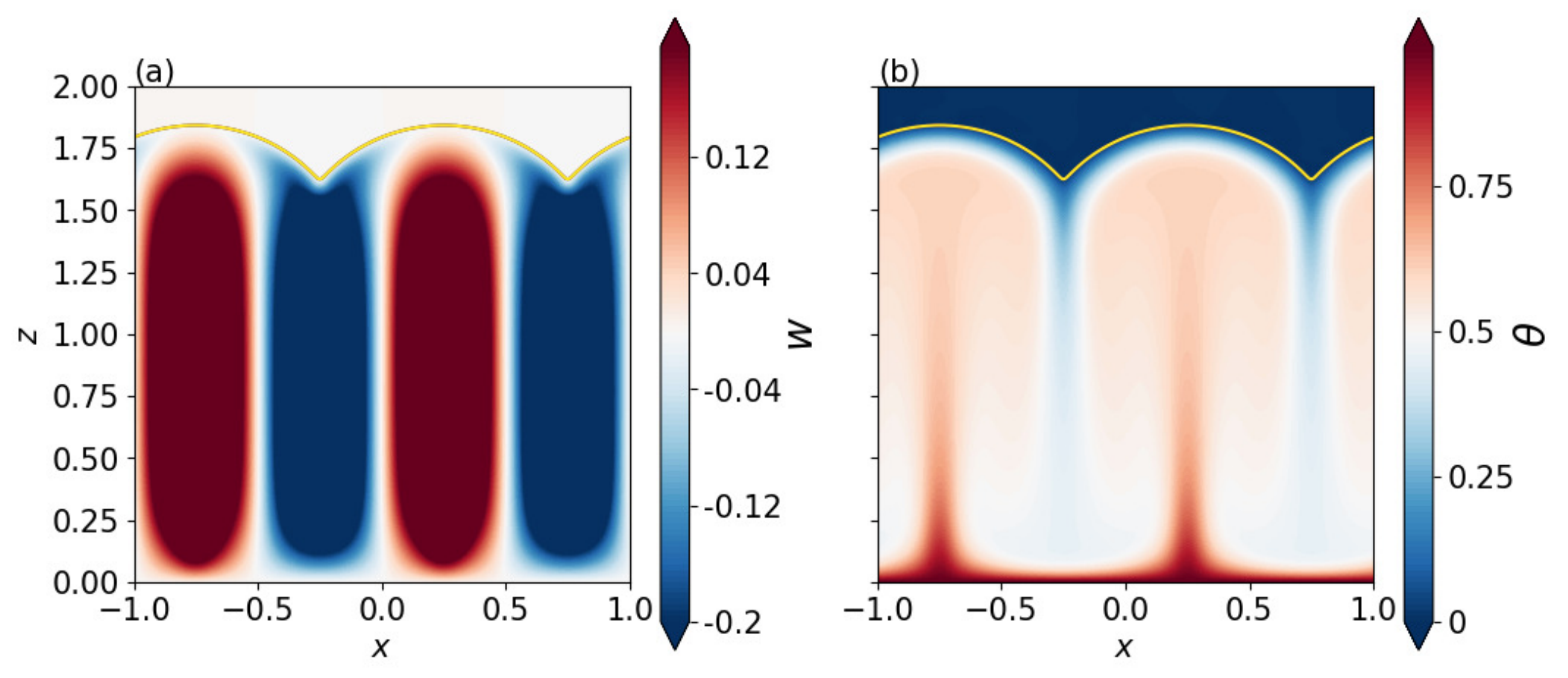}
\caption{{\label{fig:validation_favier_1} (a) The vertical velocity and (b) the temperature fields at $t=56$
for simulations at the highest resolution ($N=1024$), with $Ra=1.25\times10^5$, $Pr=1$ and $St=1$.}}
\end{figure}

\begin{figure}
\includegraphics[width=1\columnwidth]{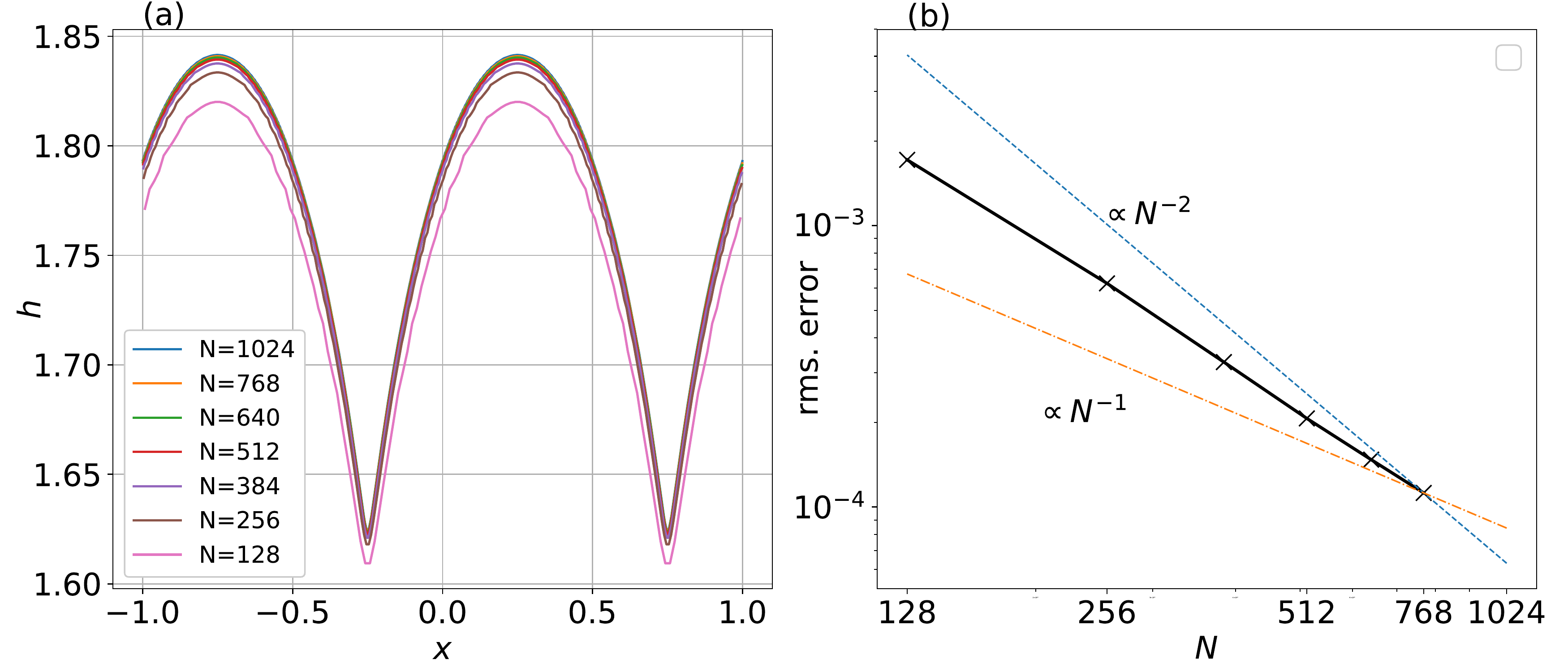}
\caption{{\label{fig:validation_favier_2} (a) The liquid height as the grid resolution is varied, and (b)
the RMS error as a function of the resolution.}}
\end{figure}

{\subsection*{Melting by convective heat transfer} \label{subsec:validation_convection}}
{The grid-dependence of the accuracy of our solution method is examined as follows.  
We use the geometry in Appendix A2 of \cite{Favier2019}, and 
$Ra=1.25\times10^5$, $Pr=1$ and $St=1$, with an initial temperature perturbation of 
\[
\theta(t=0) = 1 - z + A \text{sin}(2 \pi x) \text{sin}(\pi z).
\]
The resulting velocity and temperature
fields at $t=56$ are plotted in Fig. \ref{fig:validation_favier_1}. The location of the solid-liquid
interface is given by the liquid height $h$ from Eq. (\ref{eq:h_defn}), and is plotted as the grid resolution is varied in Fig. \ref{fig:validation_favier_2}(a).   We then use the solution at the highest grid resolution ($N=1024$) as a reference,
and linear interpolation to find the interface location at intermediate points.  The RMS error is plotted as a function of $N$ in Fig. \ref{fig:validation_favier_2}(b), showing that the error decreases as $N$ is increases, with an exponent between $1$ and $2$, as also reported by \cite{Favier2019}.}\\

\section*{Appendix B: Penalization parameter \label{sec:Appendix_penalization_parameter}}

The volume penalization method has a tunable parameter $\eta$. The
principle of the volume penalization method is to treat the solid
as a porous medium of vanishing porosity. The use of a finite value
for $\eta$ creates a velocity boundary layer of size $\left(\nu\eta\right)^{1/2}$
in the solid. \cite{Engels2015} showed that the optimal value of
$\eta$ is such that the grid spacing is comparable to the boundary
layer thickness, namely $dx\sim(\nu\eta)^{1/2}$. All of our results
are reported with the penalization parameter {$\eta=2\times10^{-3}$} (\S
\ref{sec:Numerical-Simulations}), satisfying this requirement.

\begin{figure}
\noindent \begin{centering}
\includegraphics[width=1\columnwidth]{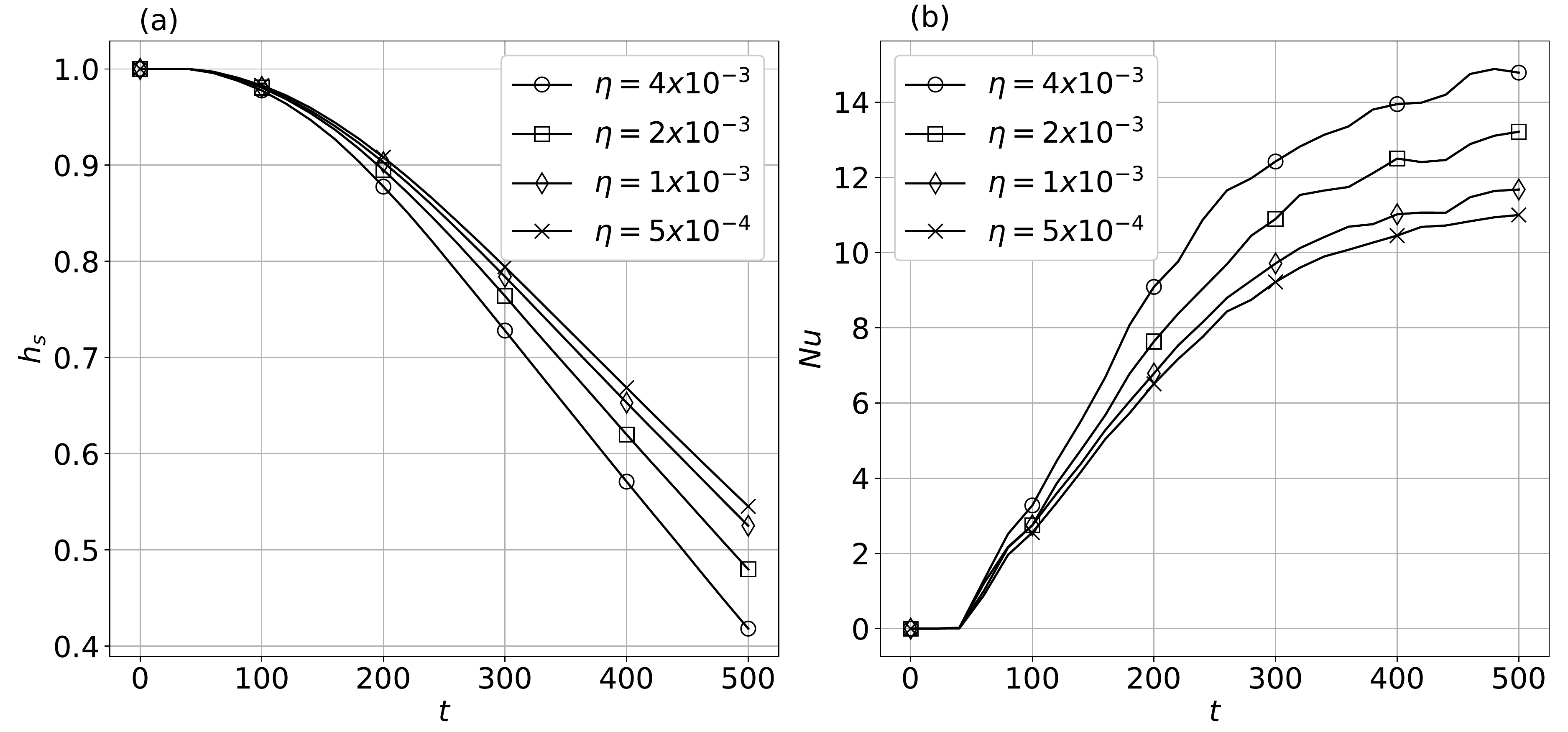}
\par\end{centering}
\caption{\label{fig:sum_chi_vs_eta} (a) The melting history and (b) the melting
Nusselt number for $E=8\times10^{-5},$ $Ra=7.8\times10^{6}$, $Pr=5$,
$St=1$, $f=0$, as $\eta$ is varied. The difference in the total
amount of solid melted changes by only about $5-10\%$ over $250$
flow units when $\eta$ is halved from {$2\times10^{-3}$} to {$10^{-3}$}.
The differences in the \emph{melting rates }are even smaller. As a
result, the Nusselt number also changes by only about $5-10\%$ as
the $\eta$ is halved. }
\end{figure}

\begin{figure}
\includegraphics[width=0.48\columnwidth]{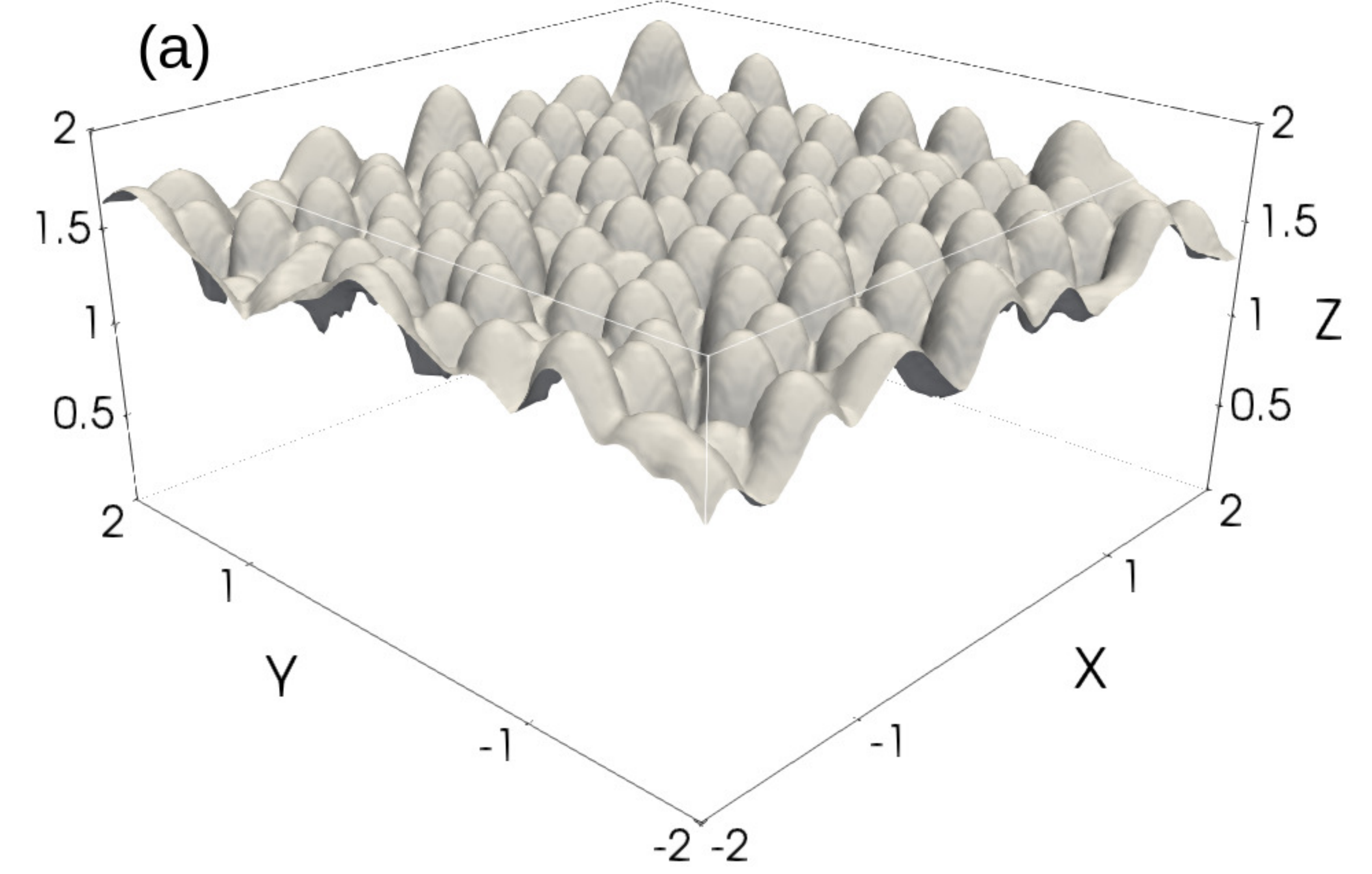}
\includegraphics[width=0.48\columnwidth]{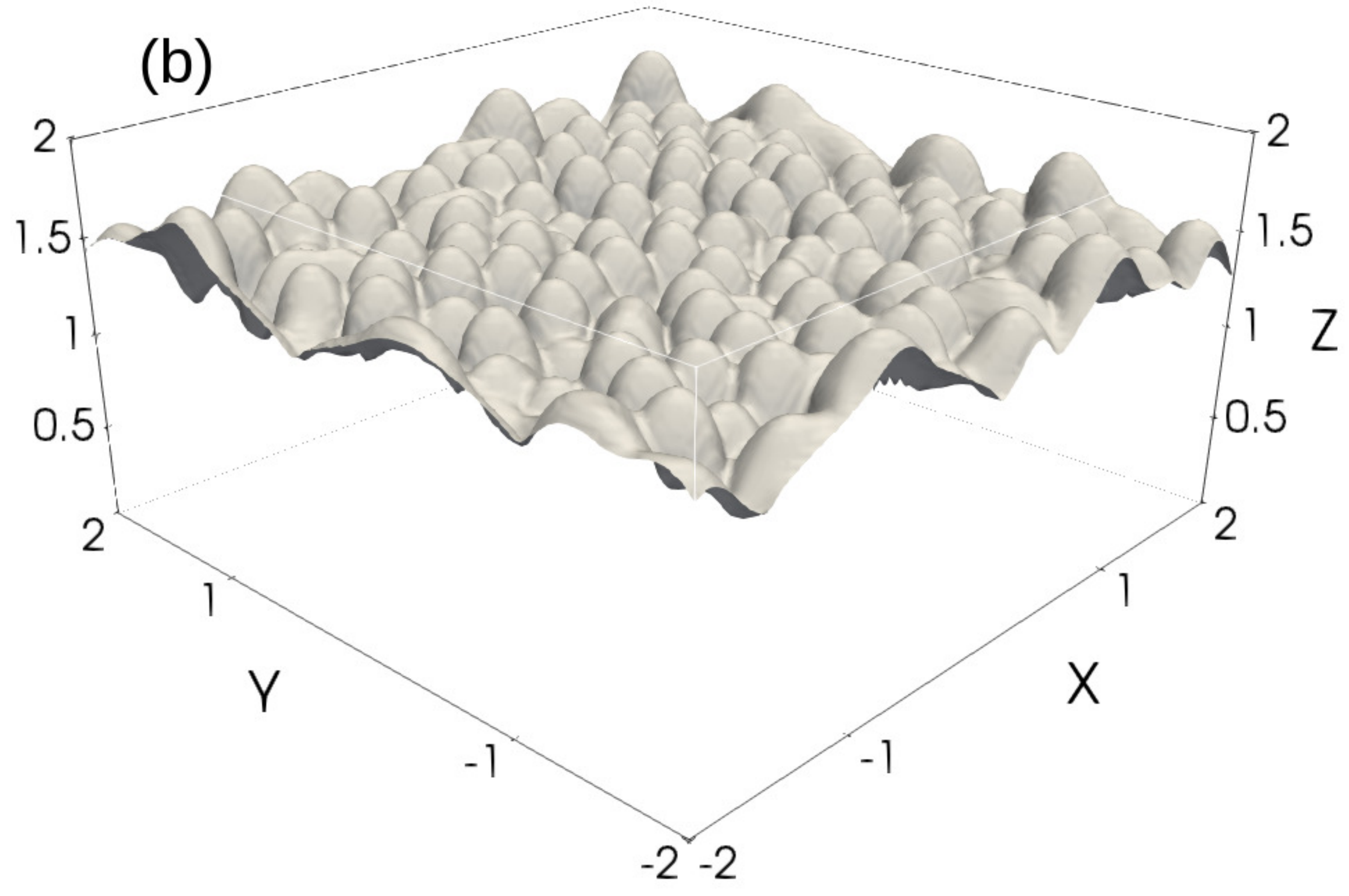}

\caption{\label{fig:contour_mask_vs_eta} Snapshots of the phase boundary at
{$t=500$} for the case $E=8\times10^{-5},$ $Ra=7.8\times10^{6}$,
$Pr=5$, $St=1$, $f=0$, with (a) {$\eta=2\times10^{-3}$} and (b) {$\eta= 10^{-3}$}.
The number and area of the voids, as well as the overall amount of
melting (noting that the figures are plotted at the same time {$t=500$}),
can be seen to be insensitive to the penalisation parameter. }
\end{figure}

In detail the melting process is influenced by the boundary layer
and hence depends on $\eta$. As seen in Fig. \ref{fig:sum_chi_vs_eta},
upon reduction of $\eta$ by a factor of $2$, the melt rate changes
by only a few percent. Therefore, the latent heat flux and the quasi-steady
balance described by Eq. \ref{eq:melting_rate} underlying the results
shown in Figs. \ref{fig:heat_balance}(b)  and
\ref{fig:Nu_sigma_h} are insensitive to the choice of $\eta$. Snapshots
of the interface shown in Fig. \ref{fig:contour_mask_vs_eta} demonstrate
the persistence of the central behavior; convective vortices etch
voids into the solid, and the number of voids are proportional to
the number of vortices. Thus, as noted in \S \ref{sec:optimal},
Nu$(t)$ and the maximal interface roughness depend on $\eta$, but
the correlation between Nu and $\sigma(h)$ shown in Figs. \ref{fig:Nu_sigma_h} do not.

\begin{figure}
\includegraphics[width=1\columnwidth]{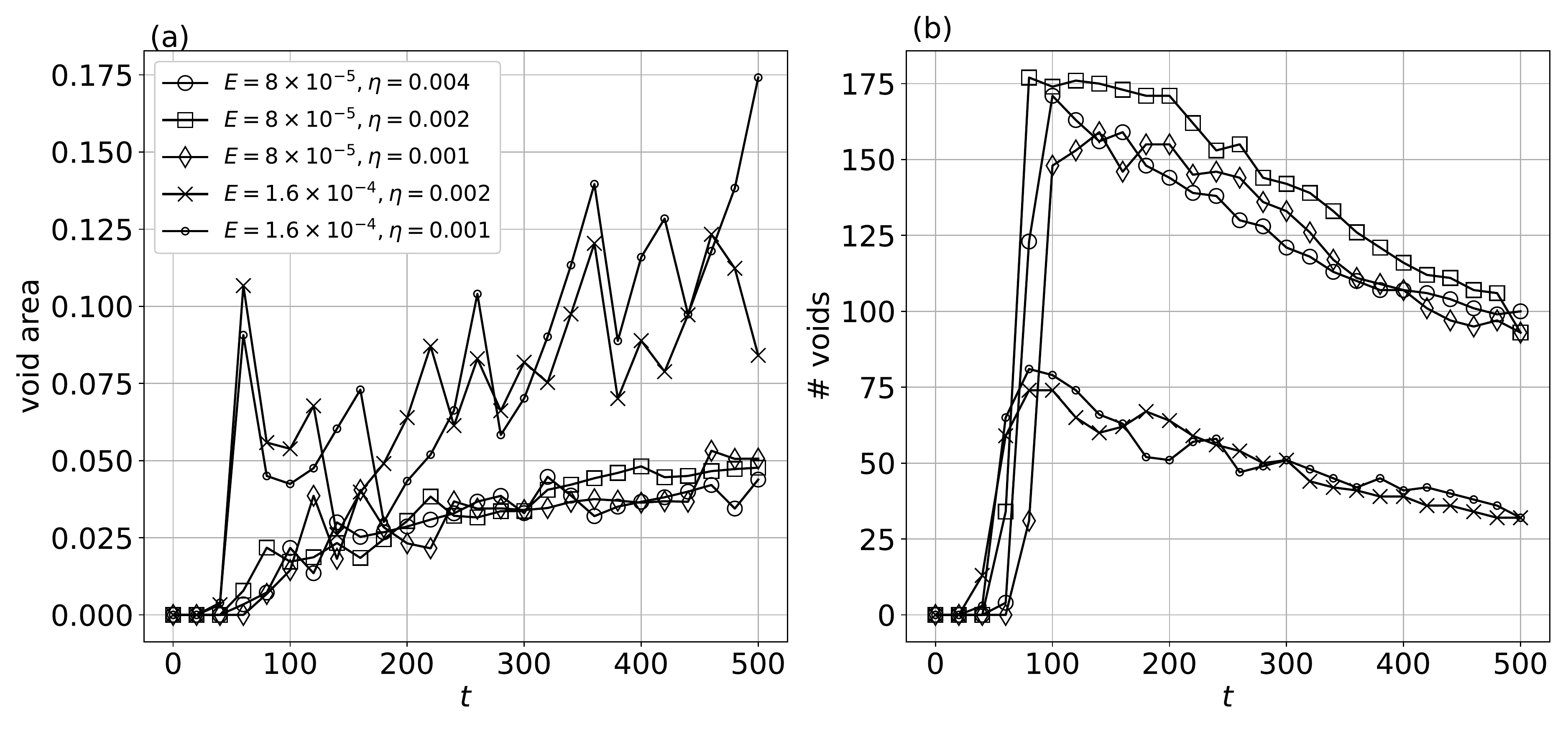}

\caption{\label{fig:nvoids_voidareas_vs_eta} The void areas and the number
of voids formed in the solid for different values of the penalisation
parameter. The number and area of the voids can be seen to be insensitive
to the penalisation parameter.}
\end{figure}

%

\end{document}